 \documentclass[authoryear,preprint,12pt]{elsarticle}

\usepackage{amssymb}
\usepackage{array}
\newcolumntype{C}[1]{>{\centering\let\newline\\\arraybackslash\hspace{0pt}}m{#1}}

 \usepackage{amsthm}
 \usepackage{amsmath}

\usepackage{booktabs}
\usepackage{tabularx}

\usepackage{multirow}
\usepackage{caption}
\usepackage{subcaption}
\usepackage{tabularx}
\usepackage{bm}

\usepackage{pgf, tikz} 
\usetikzlibrary{arrows,automata,fit}
\usetikzlibrary{shapes}

\usepackage{booktabs}
\usepackage{hyperref}

\begin{document}
\begin{frontmatter}



\title{Uncovering Drivers of EU Carbon Futures with Bayesian Networks}

\author[label1]{Jan Maciejowski}
 \author[label1]{Manuele Leonelli}
 \affiliation[label1]{organization={School of Science and Technology, IE University, Madrid}, country = {Spain}}


\begin{abstract}
The European Union Emissions Trading System (EU ETS) is a key policy tool for reducing greenhouse gas emissions and advancing toward a net-zero economy. Under this scheme, tradeable carbon credits, European Union Allowances (EUAs), are issued to large emitters, who can buy and sell them on regulated markets. We investigate the influence of financial, economic, and energy-related factors on EUA futures prices using discrete and dynamic Bayesian networks to model both contemporaneous and time-lagged dependencies. The analysis is based on daily data spanning the third and fourth ETS trading phases (2013–2025), incorporating a wide range of indicators including energy commodities, equity indices, exchange rates, and bond markets. Results reveal that EUA pricing is most influenced by energy commodities, especially coal and oil futures, and by the performance of the European energy sector. Broader market sentiment, captured through stock indices and volatility measures, affects EUA prices indirectly via changes in energy demand. The dynamic model confirms a modest next-day predictive influence from oil markets, while most other effects remain contemporaneous. These insights offer regulators, institutional investors, and firms subject to ETS compliance a clearer understanding of the interconnected forces shaping the carbon market, supporting more effective hedging, investment strategies, and policy design.
\end{abstract}



\begin{keyword}
Bayesian networks\sep Carbon Pricing \sep EU Emissions Trading System \sep EUA Futures \sep Environmental Policy
\end{keyword}

\end{frontmatter}


\section{Introduction}

The EU Emissions Trading System (ETS) is a cap-and-trade scheme designed to reduce greenhouse gas emissions across the EU, recently strengthened through the Fit for 55 package \citep{ec_ets}. Emissions-intensive sectors are required to hold European Union Allowances (EUAs) to offset their carbon output. One EUA grants the right to emit one tonne of CO$_2$. If, by the surrender deadline, a firm does not hold enough EUAs to cover its emissions, it faces financial penalties and must make up the shortfall in the following compliance cycle.

EUAs are distributed through annual auctions under a declining cap. Participants include all EU countries, Iceland, Norway, Liechtenstein, and Northern Ireland’s energy sector. Auctions occur through the European Energy Exchange in Germany, while surplus allowances may be retained for future use or traded on the secondary market, often via derivatives. Among these, futures contracts are the most liquid instruments and the focus of this study \citep{esma2024}. EUA futures enable participants to hedge price risk, with delivery typically set in December at the end of the compliance year. Certain carbon-intensive sectors receive free allowances to reduce the risk that firms relocate production to countries with less restrictive emissions regulations. These free allocations are gradually being phased out.

As the ETS directly affects energy prices and production costs, it has been central to EU political and economic discussions. Revenues from EUA auctions are intended to fund the transition to a clean economy, especially in the energy sector. However, rising EUA prices can burden energy consumers, particularly in countries with more carbon-intensive generation. This effect is often amplified by volatility in fossil fuel markets.

Understanding the drivers of EUA pricing is essential for stakeholders across policy, compliance, and finance. As an indirect form of carbon taxation, EUA revenues must be strategically allocated to support the energy transition. At the same time, compliance sectors, including energy, heat, and petroleum production, face fluctuating EUA prices and must manage risk accordingly. Financial institutions active in the market also require tools for forecasting and price discovery. A clearer understanding of EUA price drivers enables all participants to anticipate trends and volatility, improving decision-making and reducing market inefficiencies.

Prior studies have examined the economic and policy factors shaping EUA prices using various statistical approaches. We extend this work by incorporating a broader set of financial and energy-related indicators and modelling their interdependencies through discrete and dynamic Bayesian networks (BNs) \citep[e.g.][]{koller2009,murphy2002,pearl2014}. These models provide a probabilistically grounded, data-driven representation of the EUA pricing system, capturing both contemporaneous and time-lagged effects.

BNs are well suited to this setting: they represent multivariate relationships through conditional dependencies, offering interpretability and transparency that many black-box machine learning methods lack \citep{rudin2019stop}. A standard (static) BN captures associations among variables on the same day, while its dynamic counterpart introduces temporal structure to model next-day influences.

We use both static and dynamic BNs to analyze how energy, financial, and macroeconomic signals interact to shape EUA pricing. This framework reveals the interconnected structure of the variables driving carbon prices and enables the identification of key drivers, including coal and oil futures, energy sector equity performance, and broader market sentiment, alongside their contemporaneous and time-lagged effects.

The dynamic extension quantifies temporal dependencies and complements the static model, offering a richer understanding of how shocks propagate across markets. Together, the models provide actionable insights for compliance entities, financial institutions, and regulators seeking to forecast EUA prices, manage risk, or design more effective policy interventions.

\section{Literature review}

A wide range of academic and industry studies has sought to explain the drivers of EUA prices, drawing on variables from regulatory mechanisms, commodity markets, energy production, macroeconomic conditions, and financial speculation. Most of this literature relies on standard econometric techniques such as structural vector autoregressions  \citep{lovcha2021}, quantile regressions \citep{tan2017}, copula-based spillover analysis \citep{hanif2021}, and structural equation modelling \citep{wang2020}. While these approaches have yielded important insights, they are often limited in capturing the joint structure of interdependencies or the time-propagation of effects. In this section, we summarize relevant findings across key domains, highlighting evidence that motivates the inclusion of specific variables in our model and clarifying where our approach extends the current state of the art.

\subsection{Regulatory mechanisms}

The supply of EUAs is governed by a set of regulatory instruments that play a fundamental role in shaping price dynamics in the EU ETS. Among these, the Market Stability Reserve (MSR) and the Linear Reduction Factor (LRF) are the most influential mechanisms introduced to tighten the system over time and address imbalances in supply and demand. The MSR, introduced in 2019 and strengthened under the Fit for 55 package, was designed to manage the oversupply of allowances that characterized the first two trading phases. It adjusts auction volumes based on the Total Number of Allowances in Circulation (TNAC): if TNAC exceeds 833 million, 24\% of the surplus is withheld and placed in reserve; if TNAC falls below 400 million, 100 million allowances are released. Under the revised rules introduced in the Fit for 55 package, any allowances held in the MSR above the 400 million threshold are invalidated, permanently tightening supply \citep{borghesi2023, ec_ets, salvagnin2024}.

The LRF, which defines the annual rate at which the emissions cap declines, was increased from 2.2\% to 4.3\% under the Fit for 55 reform, with the goal of achieving a 55\% reduction in greenhouse gas emissions by 2030 compared to 1990 levels. These supply-side constraints are critical drivers of long-term EUA price appreciation, as they limit the availability of allowances in each compliance year \citep{ec_ets}.

Several studies have emphasized the importance of these mechanisms in supporting market stability and environmental integrity. For instance, \citet{borghesi2023} argue that the MSR was effective in raising EUA prices during the COVID-19 pandemic by absorbing excess supply; however, the mechanism might not react accordingly to longer-term shocks. Additionally, the MSR shortens the time horizon over which firms can bank allowances, increasing price volatility. Both \citet{borghesi2023} and \citet{salvagnin2024} note that regulatory uncertainty about the future configuration of these mechanisms, especially during global events or geopolitical disruptions, can lead to volatile price movements.

While these instruments are crucial in shaping expectations and long-term trends, they are not easily encoded as explanatory variables in high-frequency (e.g., daily) models. Their influence tends to be reflected in market-level responses, such as shifts in energy sector equity indices, fuel prices, or trading volume. For this reason, we do not model the MSR or LRF explicitly, but rather capture their market-mediated effects through observable variables that respond to changing regulatory constraints.

\subsection{Commodity markets}

EUA prices are closely tied to energy commodity markets, particularly coal, natural gas, and oil. Coal and gas are key fuels in the European energy mix, and their relative prices influence power producers' fuel choices, a phenomenon known as fuel switching \citep{lovcha2021}. When natural gas prices are high, producers shift toward cheaper but more carbon-intensive coal, increasing the demand for EUAs. Conversely, high coal prices create incentives to switch to cleaner alternatives, reducing EUA demand \citep{lovcha2021, tan2017}.

Empirical studies support these dynamics. \citet{lovcha2021} found significant connections between coal and gas prices and EUA price movements. Similarly, \citet{tan2017} identified a negative relationship between these commodity prices and EUA prices. The strength of these effects varies over time, reflecting shifts in energy policy, weather conditions, and fuel market fundamentals. More recent work has also confirmed the predictive importance of coal and gas in explaining EUA price variation, especially given Europe's continued reliance on fossil fuels for electricity generation \citep{salvagnin2024}.

Oil, though less directly tied to power generation, also plays an important role. Natural gas contracts are often linked to oil prices, and oil itself serves as a proxy for global economic and energy system trends \citep{lovcha2021}. Shocks to oil prices, particularly under bearish or bullish market conditions, can influence EUA prices both directly and indirectly through spillovers to financial markets and investor sentiment \citep{wang2020, zheng2021}. For example, demand-driven oil price drops can stimulate industrial activity, increasing EUA demand. Conversely, oil-related risk shocks may depress equity markets and reduce energy demand, lowering EUA prices. These effects reflect broader uncertainty: heightened risk leads investors to demand higher yields, affecting both commodity and carbon markets \citep{tan2017, zheng2021}.

\subsection{Energy markets}
The energy sector is the largest source of verified emissions under the EU ETS \citep{eea}, making its performance a critical driver of EUA prices. Changes in electricity generation costs, fuel choices, and production intensity shape allowance demand. Since electricity prices reflect two main inputs, fossil fuels and carbon permits, they are affected by EUA price movements. While the reverse effect is theoretically limited, \citet{lovcha2021} note that electricity price volatility can spill over to EUA prices by altering production incentives and sectoral demand patterns.

Electricity prices influence EUA demand both directly, by affecting energy consumption and production costs, and indirectly, by reflecting shifts in microeconomic risk. Higher electricity prices may dampen industrial output or lead to fuel switching, both of which affect the volume of emissions and corresponding demand for allowances \citep{tan2017}. These interactions underscore the need to model the energy sector’s relationship to EUA pricing as part of a broader economic and commodity network.

The ongoing transition to renewable energy adds further complexity. As the share of clean energy in the European mix rises, demand for EUAs from the power sector is expected to fall. Equity indices tracking renewable firms, such as the ERIX index, have been shown to closely mirror EUA price dynamics, with the growth of renewables associated with lower emissions and reduced allowance demand \citep{hailemariam2022,salvagnin2024}.

Paradoxically, however, lower wholesale electricity prices resulting from increased renewable penetration can spur economic activity and raise energy consumption. \citet{pena2022} showed that lower production costs are not necessarily passed on to households, due to taxes and levies, making the net effect on EUA prices ambiguous and context-dependent.

Clean energy sectors also interact with financial markets, indirectly affecting EUA pricing. \citet{wang2020} found that innovation in renewables supports equity markets and economic growth, which in turn may push EUA prices upward via increased energy demand. Further, \citet{hanif2021} identified the EUA market as a net spillover recipient from clean energy indices, including the S\&P Global Clean Energy Index and the NYSE Bloomberg Global WIND Index. These findings reinforce the need to account for clean energy as an evolving and influential component in EUA price formation.

\subsection{European economic conditions}

Macroeconomic and financial indicators, including equity markets, bond yields, exchange rates, and commodity prices, shape EUA pricing by influencing expectations about energy demand and industrial activity. \citet{lovcha2021} found a connection between EUA prices and national stock indices that varies across the economic cycle. Economic growth typically leads to more production and emissions, raising EUA demand, while downturns reduce consumption and dampen allowance prices.

Eurozone bond yields offer further insight into market expectations. Shifts in these yields can trigger changes in EUA pricing by altering the outlook for growth and investment \citep{salvagnin2024}. \citet{tan2017} noted that adverse macroeconomic conditions may raise financing costs, constrain investment, and reduce emissions, indirectly lowering EUA demand. Currency fluctuations, too, convey global risk sentiment; sharp exchange rate movements or uncertainty can spill into EUA markets, especially during periods of speculation or instability \citep{salvagnin2024}. High-yield bond spreads, often viewed as indicators of credit risk and market stress, have been linked to EUA price volatility \citep{tan2017}.

The relative importance of these factors has shifted over time. \citet{salvagnin2024} observed that financial-market variables became more prominent drivers of EUA prices in Phase 4 of the ETS, shaped by the COVID-19 pandemic and the energy crisis linked to the Russo-Ukrainian war.

Bonds, both green and conventional, offer further explanatory power. Green bonds fund environmental investments and may signal shifts in expectations around emissions trajectories. \citet{leitao2021} found that the S\&P Green Bond Index had a positive and significant short-term effect on EUA prices across volatility regimes. Similar results were observed for the Sol Green Index. These findings challenge earlier assumptions that clean energy growth would reduce EUA demand \citep{koch2014}, possibly due to rising carbon price expectations or delays in the fossil-to-renewable transition. \citet{leitao2021} also suggest that switching costs and infrastructure inertia may sustain fossil fuel use, and EUA demand, in the near term.

In contrast, conventional bonds show a negative relationship with EUA prices, significant only in volatile markets \citep{leitao2021}. Concerning non-investment grade bonds, higher bond yields, reflecting greater economic risk, can suppress industrial output and energy use, leading to lower EUA demand \citep{tan2017}.

\subsection{Role of speculation}
The role of speculation in EUA pricing and volatility has attracted sustained attention, especially in EU policy discussions. Concerns stem from the possibility that speculative activity distorts the “true” price of carbon allowances, an issue made more critical given that EUA prices directly affect electricity pricing \citep{lovcha2021}. The EUA market exhibits characteristics conducive to speculation, including high liquidity, standardisation, low transaction costs, and price volatility \citep{icis2020}.

Since EUAs were designated as financial instruments, the European Securities and Markets Authority (ESMA) has overseen market regulation. Trading is highly concentrated: 97\% of secondary volume occurs on the Intercontinental Exchange (ICE), and 90\% of auctioned allowances are acquired by just ten financial intermediaries \citep{esma2024,icis2020}. Many ETS operators rely on these institutions for efficiency and expertise. While compliant firms tend to hold long-term positions, banks and investment firms often take short-term ones. In 2023, ESMA reported that 73\% of secondary market volume involved financial institutions, with 39\% of trades linked to low-risk spread strategies. The prevalence of algorithmic and high-frequency trading is not necessarily harmful, as these strategies are not inherently speculative.

Speculation plays a limited role in long-term price trends but is a major driver of short-term variability. \citet{lovcha2021} found that speculative behaviour explains much of the short-term price dynamics, particularly relevant for market participants exposed to high-frequency movements.

Further evidence comes from bubble analysis. \citet{terranova2024} identified seven speculative bubbles in EUA prices between 2017 and 2022, typically triggered by regulatory announcements. These bubbles lasted an average of 22 days, with speculative dynamics present on as much as 10\% of trading days during the sample period, raising concerns about the potential distortion of market signals.

In sum, while long-term EUA price trends are driven by fundamentals, speculation introduces short-term volatility that can obscure underlying dynamics. Recognising this influence remains important when interpreting EUA market behaviour.

\subsection{Summary and outlook}

In sum, the literature highlights a wide range of economic, energy, and financial factors influencing EUA prices, often through complex and time-dependent mechanisms. While previous studies have established the relevance of these drivers, most rely on methods that treat them in isolation or fail to model their interdependencies. A glossary of key terms and a summary of the main reviewed works are provided in the appendix (Tables~\ref{tab:termglossary} and~\ref{tab:litrevworks}, respectively). Building on these insights, the remainder of this paper introduces a probabilistic framework based on discrete and dynamic BNs, which allows for joint modeling of contemporaneous and lagged effects across variables influencing the carbon market.

\section{Methodology}

To model the relationships among the variables influencing EUA futures prices, we develop two probabilistic graphical models. The first is a standard discrete Bayesian network (BN) that captures contemporaneous dependencies among variables. Building on its structure, we then construct a dynamic Bayesian network (DBN) to incorporate time-lagged relationships and explore how changes in one variable affect others across successive time steps. This extension enables a more complete view of both intra- and inter-temporal effects, particularly relevant in the context of financial and energy market interactions. In this section, we describe the data used in the analysis, the pre-processing steps undertaken, and the modeling framework used to uncover the structure and strength of dependencies across the system.

\subsection{Data}

\begin{figure}
\centering
\includegraphics[width=\linewidth]{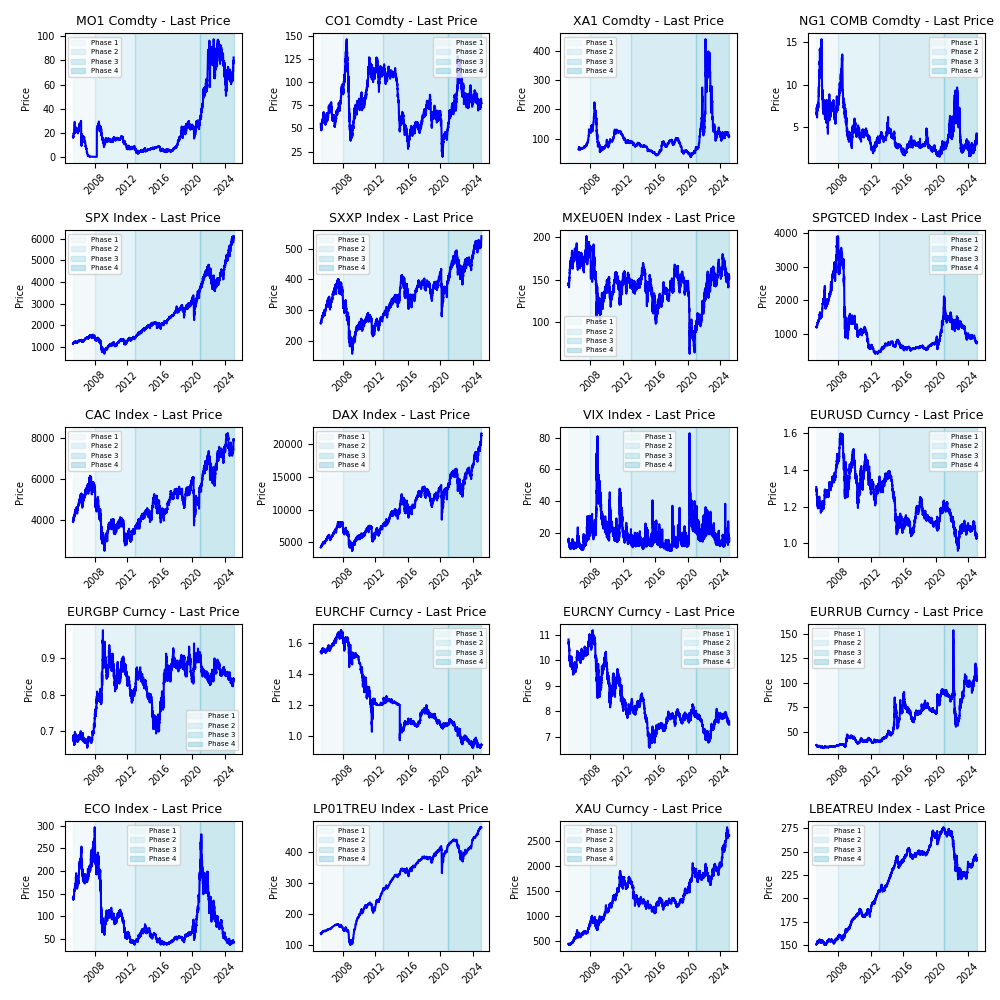}
\caption{Daily price trajectories for all variables included in the analysis, spanning Phases 3 to 4 of the EU ETS. Shaded regions denote the different compliance phases. Prices are shown on raw scales and represent the last available value for each trading day.}
\label{fig:timegraphs}
\end{figure}

Guided by the literature review, we assembled a set of 20 variables to model the factors influencing EUA futures prices. These include energy commodities, stock indices, exchange rates, bond indices, volatility measures, and clean energy indicators. The dataset spans the third and fourth phases of the EU ETS, covering daily observations from January 3, 2013, to January 31, 2025. All data were retrieved from Bloomberg to ensure consistency and reliability. A time series overview of the selected variables is provided in Figure~\ref{fig:timegraphs}, and full variable names and Bloomberg tickers are listed in Table~\ref{tab:variables}.

\begin{table}
\centering
\footnotesize
\resizebox{\textwidth}{!}{
\begin{tabularx}{\textwidth}{l l X l}
\toprule
\textbf{Category} & & \textbf{Variable Name} & \textbf{Bloomberg Ticker} \\
\midrule
Dependent Variable & & ICE European Union Allowance Futures (1st month) & MO1 Comdty \\
\midrule
Energy Commodities & & ICE Brent Crude Oil Futures (1st month) & CO1 Comdty \\
 & & ICE API2 Rotterdam Coal Futures (1st month) & XA1 Comdty \\
 & & Natural Gas Futures (1st month, composite pricing) & NG1 COMB Comdty \\
\midrule
Stock Indexes & & S\&P 500 Index & SPX Index \\
 & & STOXX Europe 600 Price Index & SXXP Index \\
 & & French CAC 40 & CAC Index \\
 & & German DAX & DAX Index \\
\midrule
Volatility Measures & & Chicago Board Options Exchange Volatility Index & VIX Index \\
 & & Gold Spot Rate & XAU Curncy \\
\midrule
Energy Indexes & & MSCI Europe Energy Index & MXEU0EN Index \\
 & & S\&P Global Clean Energy Index & SPGTCED Index \\
 & & WilderHill Clean Energy Index & ECO Index \\
\midrule
Bond Indexes & & Bloomberg Pan-European High Yield Total Return Index & LP01TREU Index \\
 & & Bloomberg EuroAgg Total Return Index Value Unhedged EUR & LBEATREU Index \\
\midrule
FX Rates & & Euro to U.S. Dollar Currency Exchange Rate & EURUSD Curncy \\
 & & Euro to British Pound Sterling Currency Exchange Rate & EURGBP Curncy \\
 & & Euro to Swiss Franc Currency Exchange Rate & EURCHF Curncy \\
 & & Euro to Chinese Renminbi Currency Exchange Rate & EURCNY Curncy \\
 & & Euro to Russian Ruble Currency Exchange Rate & EURRUB Curncy \\
\bottomrule
\end{tabularx}}
\caption{List of variables included in the analysis, grouped by category, with their corresponding Bloomberg tickers.}
\label{tab:variables}
\end{table}

Energy commodities were included due to their central role in shaping emissions and compliance strategies under the ETS. Coal and natural gas remain core fuels in European electricity generation, and their price dynamics influence the demand for carbon allowances through fuel-switching behavior. Brent crude oil, while less directly tied to electricity production, serves as a key benchmark for global energy markets and often co-moves with gas prices due to contractual linkages. Together, these variables provide insight into the supply-side conditions of the energy market that influence EUA pricing.

To complement commodity-level data, we include equity indices that reflect both traditional and clean energy sectors. The MSCI Europe Energy Index captures the performance of major European companies subject to EUA surrender obligations, while the S\&P Global Clean Energy and WilderHill ECO indices track renewable energy investments. These clean energy proxies are particularly relevant in the context of decarbonization, as rapid growth in renewables is expected to reduce EUA demand over time, albeit with transitional effects that may increase short-term volatility.

Financial and macroeconomic conditions are captured through a combination of stock indices, currency pairs, bond yields, and volatility indicators. Equity indices, including the DAX, CAC 40, STOXX Europe 600, and S\&P 500, serve as forward-looking indicators of market sentiment and industrial activity across European and global contexts. Five Euro exchange rates reflect the economic performance and risk sentiment of major trading partners. The VIX index measures implied volatility in U.S. equity markets, acting as a global barometer of uncertainty, while the gold spot rate represents a traditional safe-haven asset.

Bond indices provide additional macro-financial context. The Bloomberg Pan-European High Yield Index represents lower-rated corporate credit and is often used to gauge market stress, while the Bloomberg EuroAgg Index aggregates higher-quality government and investment-grade corporate bonds. Movements in these indices can reflect changing borrowing conditions, credit risk perceptions, and investor risk appetite, all of which may influence energy investment and EUA demand.

\subsection{Data pre-processing}

Before modeling, the raw data were transformed into standardized daily log returns to address differences in scale and ensure comparability across variables. Missing values, typically due to market holidays, were imputed using the most recent available price, a common practice in financial time series settings.

To isolate contemporaneous relationships in the static BN and reduce noise from autocorrelation and volatility clustering, we applied AR-GARCH filtering to each time series \citep[e.g.][]{ruppert2015}. For the dynamic BN, we retained temporal autocorrelation and applied only GARCH filtering to remove conditional heteroskedasticity, allowing the network to capture both auto- and cross-dependencies. In both cases, standardized residuals were used as inputs to the structure learning algorithms. This procedure is widely adopted in financial applications involving probabilistic models \citep[e.g.][]{mcneil2000estimation}. Model orders were selected using the Bayesian Information Criterion (BIC), and all estimations were performed using the \texttt{arch} Python library \citep{sheppard2023arch}.

Exploratory data analysis and GARCH model specifications are provided in~\ref{appendix:eda} and~\ref{appendix:archgarch}, respectively. Code, trained models, and visualizations are publicly available at: \url{https://github.com/Majon911/CapstoneETSPublic}.

To facilitate BN modeling, all variables were discretized using quantile binning. While BNs can be extended to continuous domains, learning such models is computationally intensive and often yields limited interpretability. In contrast, discretized models produce conditional probability tables that enable clearer insights into variable interactions. Following \citet{nojavan2017}, we divided each variable’s return distribution into three categories: “Low” (below the 33rd percentile), “Neutral” (between the 33rd and 66th percentiles), and “High” (above the 66th percentile). This partitioning avoids heavily biased priors \citep{beuzen2018} and improves the model’s sensitivity to shifts in conditional dependencies. The binned variables were then used as input for both the discrete BN and DBN structure estimation.

\subsection{Discrete Bayesian network}

\subsubsection{Basic principles of Bayesian networks}

A Bayesian network (BN) is a probabilistic graphical model that represents the joint distribution of a set of random variables through a directed acyclic graph (DAG) \citep[e.g.][]{koller2009,murphy2002,pearl2014}. In this graph, each node corresponds to a variable, and each directed edge encodes a direct dependency between the connected variables. The absence of an edge implies conditional independence, formalized through the d-separation criterion \citep{pearl2014}.

Given a set of $n$ discrete variables $X_1, \dots, X_n$, a BN defines the joint probability distribution as a product of conditional probabilities:

\begin{equation}
P(X_1, X_2, \dots, X_n) = \prod_{i=1}^n P(X_i \mid \text{Parents}(X_i)),
\end{equation}

\noindent
where $\text{Parents}(X_i)$ denotes the set of immediate predecessors of $X_i$ in the DAG. This factorization allows BNs to represent high-dimensional distributions in a compact and interpretable way, especially when many conditional independencies are present.

The strength of BNs lies in their ability to combine probabilistic reasoning with a visual representation of the relationships between variables. This makes them particularly well suited for applications where both inference and interpretability are essential. As a result, BNs have been widely applied in fields such as environmental modeling, energy forecasting, and climate risk analysis \citep[e.g.][]{borunda2016bayesian,kaikkonen2020bayesian,machado2023risk,moe2020increased}. Their capacity to model uncertainty and uncover causal or structural dependencies makes them a valuable tool in understanding complex, interrelated systems such as carbon and energy markets.

\subsubsection{Learning Bayesian networks}

Although BNs can be specified manually using expert knowledge \citep{barons2022balancing,nyberg2022bard}, in this study we adopt a fully data-driven approach. Both the structure of the network (i.e., the DAG) and the associated probabilities are learned from data.

We explored a range of structure learning algorithms, including constraint-based, score-based, and hybrid approaches, as well as multiple scoring functions \citep[see e.g.][]{kitson2023,scutari2019}. Ultimately, we selected a score-based Tabu Search guided by the Bayesian Dirichlet equivalent uniform (BDeu) score \citep{heckerman1995learning}, as it provided the best balance between fit, parsimony, and alignment with prior economic understanding. Tabu Search has been shown to perform efficiently and accurately across a wide range of domains \citep{scutari2019}.

The BDeu score evaluates how well a given graph structure $G$ explains the data $D$, incorporating both likelihood and prior structure beliefs. It is defined as:

\begin{equation}
\text{Score}_{\text{BDeu}}(G \mid D) = \log P(G) + \sum_{i=1}^n \text{score}(X_i, \text{Pa}_i),
\end{equation}

\begin{equation}
\text{score}(X_i, \text{Pa}_i) = \sum_{j=1}^{q_i} \left[ \log \frac{\Gamma(\frac{N'}{q_i})}{\Gamma(N_{ij} + \frac{N'}{q_i})} + \sum_{k=1}^{r_i} \log \frac{\Gamma(N_{ijk} + \frac{N'}{r_i q_i})}{\Gamma(\frac{N'}{r_i q_i})} \right],
\end{equation}

\noindent where $X_i$ is a node in the network with $r_i$ possible states and $q_i$ parent configurations. $N_{ijk}$ is the count of observations where $X_i$ is in state $k$ given parent configuration $j$, and $N_{ij} = \sum_k N_{ijk}$. The hyperparameter $N'$ represents the equivalent sample size, which smooths the estimates and encodes prior strength \citep{heckerman1995learning}. In our implementation, we set $N' = 10$ following standard recommendations \citep{kitson2023}.

To ensure robustness and reduce sensitivity to specific data realizations, we adopted a non-parametric bootstrap strategy: the structure learning procedure was repeated over 200 bootstrapped datasets, and only edges that appeared in at least 50\% of the resulting graphs were retained in the final network. This ensemble approach improves structural stability and mitigates overfitting \citep{scutari2013}.

After learning the graph structure, we estimated the conditional probabilities using maximum likelihood estimation (MLE). For a given variable $X_i$, in state $x_k$, and parent configuration $x_j$, the conditional probability is computed as:

\begin{equation}
\hat{P}(X_i = x_k \mid \text{Pa}(X_i) = x_j) = \frac{N_{ijk}}{\sum_k N_{ijk}},
\end{equation}

\noindent where $N_{ijk}$ denotes the number of instances where $X_i = x_k$ and its parents are in configuration $x_j$. MLE provides consistent and interpretable estimates when sufficient data are available and is widely used in BN learning \citep{heckerman1995learning}.

All models were implemented in Python using the \texttt{pgmpy} library \citep{ankan2024}. The final structure and associated conditional probability tables are available through the \texttt{bnRep} R package \citep{leonelli2025bnrep}, providing an accessible resource for future research and replication.

\subsubsection{Analyses based on the Bayesian network model}

Having constructed and fitted the static BN, we investigate its structure and implications using a comprehensive suite of inference and sensitivity analysis techniques. These methods enable a detailed evaluation of the model’s internal logic, highlight key relationships, and quantify the relevance and robustness of variables influencing EUA futures prices. Our approach combines established tools with several recent methodological advances.

We begin by examining the equivalence class of the learned DAG to identify which connections can be interpreted as causal and which remain ambiguous. The associated partially directed acyclic graph (PDAG) highlights edges whose orientation is reliably supported by the data, helping avoid overconfident conclusions about causal direction \citep{chickering2002learning}.

To understand how market configurations relate to EUA outcomes, we apply two complementary inference procedures. First, we compute the most probable explanation (MPE) for each outcome of the EUA node. The MPE identifies the most likely joint configuration of the remaining variables conditional on a high, neutral, or low EUA return, offering insight into plausible market scenarios for each state of the target variable \citep{kwisthout2011most}. Second, we perform evidence propagation: each variable is fixed to a specific state, and the resulting change in the distribution of the EUA node is recorded. This enables the comparison of directional effects and highlights variables with the strongest influence on carbon pricing.

We then assess the marginal importance of each variable using three complementary measures:

\begin{itemize}
\item \textbf{Mutual information} quantifies the overall statistical dependence between each node and the EUA futures variable \citep[computed using the \texttt{bnmonitor} R package][]{leonelli2023sensitivity}.
\item \textbf{Variance component Sobol indices} measure how much observing each variable reduces the uncertainty in EUA pricing, capturing its individual contribution to the overall variability in the target \citep{ballester2022computing}.
\item \textbf{Diameter-based arc strength} quantifies the maximum influence a parent node can exert on its child in terms of total variation distance across the conditional probability table \citep{leonelli2025diameter}.
\end{itemize}

Finally, we explore local parameter sensitivity using tornado plots\footnote{Created using the GeNie software.}. These visualize the parameters with the greatest impact on the marginal distribution of the target variable, as measured by their sensitivity values. Each sensitivity value quantifies how much the output probability can change in response to structured perturbations of a single entry in the conditional probability tables \citep{ballester2023yodo}.

Taken together, these analyses provide a layered understanding of the BN’s probabilistic structure. They allow us to assess not only which variables matter most, but also how robust these conclusions are to uncertainty in model parameters, thereby offering actionable insight for risk management, market regulation, and investment strategies in the context of the EU ETS.

\subsection{Dynamic Bayesian network model}

To assess temporal dependencies among the variables influencing EUA pricing, we extend our analysis using a discrete dynamic BN (DBN). A DBN models the evolution of a system over time by introducing directed connections between variables at consecutive time steps. We adopt a two-time-slice BN representation under standard first-order Markov and stationarity assumptions: the distribution at time \( T+1 \) depends only on variables at time \( T \), and the conditional structure remains invariant across time \citep{koller2009, murphy2002}. This assumption is particularly suitable in financial contexts, where daily asset returns are typically modeled as conditionally independent of the past beyond one lag.

The dynamic model consists of two components: a static BN describing the joint distribution of variables at time \( T \),
\begin{equation}
P(X^T) = \prod_{i=1}^{n} P(X_i^T \mid \text{Pa}(X_i^T)),
\end{equation}
and a transition model that captures temporal effects through inter-slice dependencies:
\begin{equation}
P(X^{T+1} \mid X^T) = \prod_{j=1}^{n} P(X_j^{T+1} \mid \text{Pa}(X_j^{T+1})),
\end{equation}
where \( \text{Pa}(X_j^{T+1}) \subseteq X^T \). The intra-slice structure at time 
$T$ is fixed based on the static BN. To isolate temporal effects, we only learn inter-slice edges from variables at time 
$T$ to those at $T+1$, using the same score-based procedure as for the static model and excluding any within-slice connections at $T+1$. 

The resulting DBN allows us to evaluate how shocks to variables at time \( T \) affect the predicted distribution of EUA returns at time \( T+1 \), complementing the contemporaneous insights provided by the static BN.

\section{Results}

\subsection{Structure and interpretation of the learned Bayesian network}

\begin{figure}
\centering
\includegraphics[width=1\textwidth]{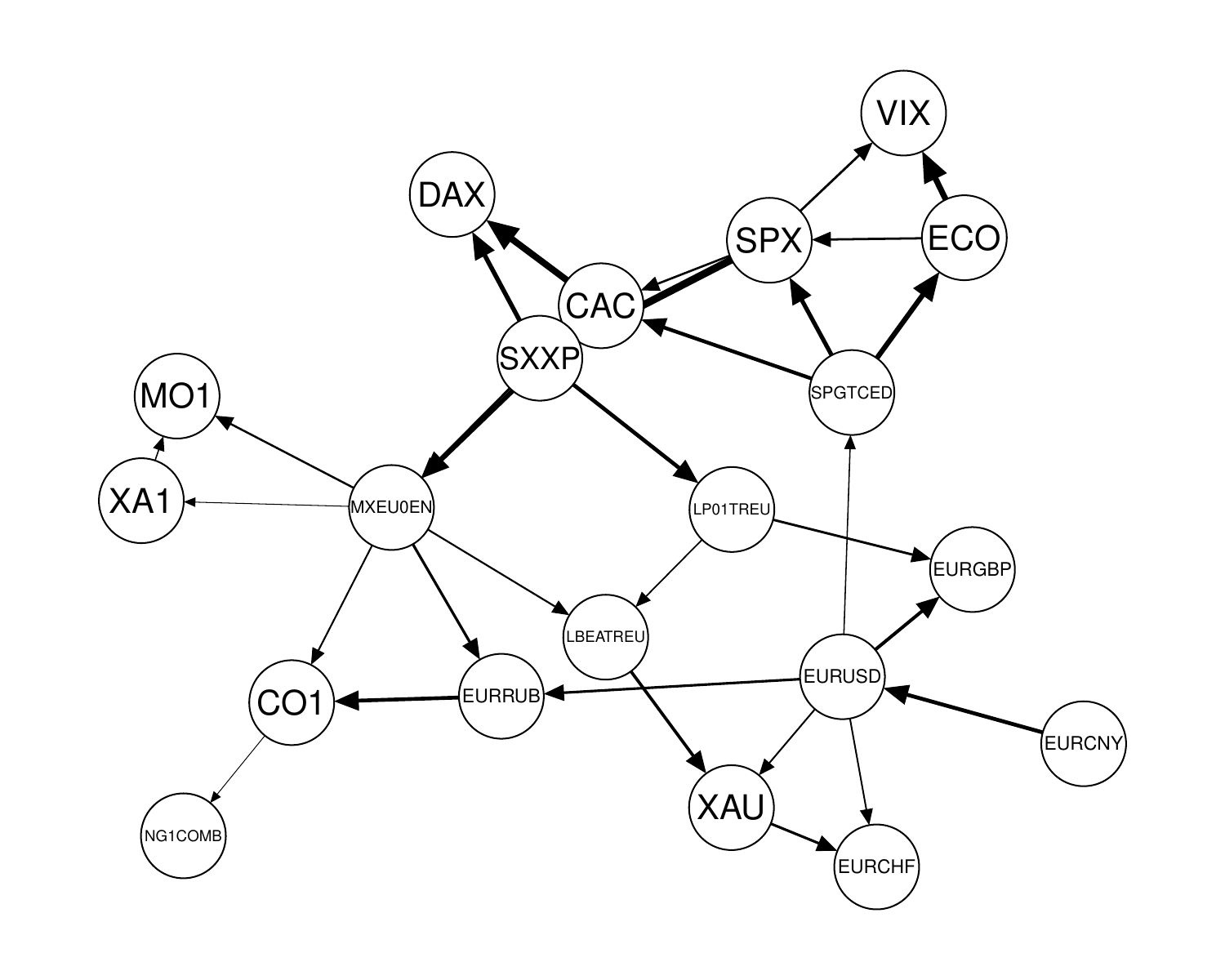}
\caption{Structure of the learned BN. Edge widths are proportional to arc strength  \citep{leonelli2025diameter}. The layout and visualization follow the \texttt{qgraph} R package \citep{epskamp2012qgraph}.}
\label{fig:bayesnet}
\end{figure}

The structure of the learned BN is displayed in Figure~\ref{fig:bayesnet}, capturing the probabilistic dependencies among the 20 variables influencing EUA futures prices. The network reveals several well-defined clusters that align with known relationships in financial and energy systems, offering an interpretable summary of key interactions.

One clear grouping includes major equity indices. The S\&P 500 is directly connected to the VIX volatility index and to the ECO clean energy index, forming a cluster representing U.S. financial market conditions. The STOXX Europe 600, CAC 40, and DAX indices form a parallel European cluster, which also shares links with the S\&P 500, highlighting the interdependence between U.S. and European equity markets. These connections are consistent with international spillovers in investor sentiment and macroeconomic expectations.

A second cluster is formed around the energy sector. The MSCI Europe Energy Index (MXEU0EN) appears as a central node, with outgoing edges to coal (XA1) and Brent oil (CO1) futures. This reflects the strong linkage between energy sector performance and fossil fuel prices. The EUA futures variable (MO1) is connected to both the MSCI energy index and coal prices, underlining the sector’s central role in driving carbon pricing dynamics.

Further groupings emerge in the foreign exchange domain. The U.S. Dollar (EURUSD) influences multiple exchange rates, including the British Pound (EURGBP), Swiss Franc (EURCHF), and Russian Ruble (EURRUB), as well as the gold spot price (XAU). This reflects the dominant role of the dollar in global financial conditions. The Russian Ruble, in turn, affects Brent oil prices, consistent with Russia’s major role in global energy exports.

Finally, a smaller cluster connects bond indices to equity markets. The MSCI Europe Energy Index shows a directed influence on the Bloomberg EuroAgg Investment Grade Index (LBEATREU), suggesting that changes in energy sector equity valuations may help explain shifts in credit market dynamics. High-yield bonds (LP01TREU) are linked to other financial market variables but show a less central role in the network.

While the learned network reveals a rich structure of dependencies, not all edge directions can be interpreted causally. Based on the equivalence class of the final model, a number of arcs are supported as directed: for example, from the MSCI Europe Energy Index to Brent oil and coal futures, from the S\&P 500 to the STOXX Europe 600 and the VIX, and from the energy index to the EUA futures node. These links are consistent with expected market dynamics. In contrast, edges among coal, energy equities, and EUA futures remain undirected, reflecting close co-movement patterns where the data do not distinguish cause from effect.


\subsection{Probabilistic inference and key drivers}

We begin by analyzing the most probable joint configurations associated with different EUA outcomes. Table~\ref{tab:mpe_table} reports the most probable explanation (MPE) under each state of the EUA node. High EUA returns tend to co-occur with elevated energy commodity prices, strong equity performance, and subdued volatility. In contrast, low EUA returns are most likely under adverse market conditions, depreciated equity indices, strong currencies, and high VIX levels. The neutral state is characterized by mixed signals, with some clean energy indices remaining high while fossil fuels revert to baseline levels.

\begin{table}
\centering
\renewcommand{\arraystretch}{1.2}
\footnotesize
\scalebox{0.78}{
\begin{tabular}{lccc}
\toprule
\textbf{Instrument} & \textbf{MO1 = High} & \textbf{MO1 = Low} & \textbf{MO1 = Neutral} \\
\midrule
CAC Index         & High    & Low     & Neutral \\
CO1 Comdty        & High    & Low     & Neutral \\
DAX Index         & High    & Low     & Neutral \\
ECO Index         & High    & Low     & High    \\
EURCHF Curncy     & Low     & High    & High    \\
EURCNY Curncy     & Low     & High    & High    \\
EURGBP Curncy     & Low     & High    & High    \\
EURRUB Curncy     & Low     & High    & High    \\
EURUSD Curncy     & Low     & High    & High    \\
LBEATREU Index    & Low     & Low     & Low     \\
LP01TREU Index    & High    & Low     & Low     \\
MXEU0EN Index     & High    & Low     & Neutral \\
NG1 COMB Comdty   & High    & Low     & High    \\
SPGTCED Index     & High    & Low     & High    \\
SPX Index         & High    & Low     & High    \\
SXXP Index        & High    & Low     & Neutral \\
VIX Index         & Low     & High    & Low     \\
XA1 Comdty        & High    & Low     & Neutral \\
XAU Curncy        & Low     & High    & High    \\
\bottomrule
\end{tabular}}
\caption{Most probable explanation (MPE) for each level of EUA return (MO1 Comdty) based on the BN.}
\label{tab:mpe_table}
\end{table}

We next assess the impact of perturbing individual variables on EUA pricing. For each node, we fix its value to a high or low state and observe the resulting distribution of the EUA node. As shown in Table~\ref{tab:evidtop10}, the largest shifts occur for coal futures (XA1) and the MSCI Europe Energy Index (MXEU0EN). Setting these to “high” increases the probability of a high EUA return from the baseline of 33\% to 43\% and 40\%, respectively; when set to “low,” this probability drops to 25\% and 27\%. Broad equity indices such as the STOXX Europe 600, CAC 40, and DAX also show directional effects, with consistent increases in EUA prices under favorable market conditions. Energy-linked indicators (Brent oil, clean energy indices) and high-yield bonds exhibit moderate influence, reinforcing the interconnected role of macroeconomic sentiment and energy fundamentals in shaping EUA price expectations.

\begin{table}
\centering
\footnotesize
\scalebox{0.78}{
\renewcommand{\arraystretch}{1.2}
\begin{tabular}{l|ccc|ccc}
\toprule
\multirow{2}{*}{\textbf{Node}} &
\multicolumn{3}{c|}{\textbf{Node = High}} &
\multicolumn{3}{c}{\textbf{Node = Low}} \\
 & High & Neutral & Low & High & Neutral & Low \\
\midrule
Coal Futures (XA1) & 43\% & 32\% & 25\% & 25\% & 32\% & 43\% \\
MSCI Energy (MXEU0EN) & 40\% & 33\% & 27\% & 27\% & 32\% & 42\% \\
STOXX Europe 600 & 37\% & 33\% & 30\% & 30\% & 33\% & 37\% \\
CAC 40 & 37\% & 33\% & 30\% & 30\% & 33\% & 37\% \\
DAX & 36\% & 33\% & 30\% & 30\% & 33\% & 37\% \\
Brent Oil (CO1) & 36\% & 33\% & 30\% & 31\% & 33\% & 36\% \\
S\&P 500 (SPX) & 35\% & 33\% & 32\% & 32\% & 33\% & 35\% \\
S\&P Clean Energy & 35\% & 33\% & 32\% & 32\% & 33\% & 35\% \\
High-Yield Bonds & 35\% & 33\% & 32\% & 32\% & 33\% & 35\% \\
ECO Index & 34\% & 33\% & 32\% & 32\% & 33\% & 35\% \\
\bottomrule
\end{tabular}}
\caption{Top 10 variables influencing EUA futures price distribution. Each row shows the conditional probability of a high, neutral, or low EUA return, given a high or low state of the influencing node. Values are rounded to nearest percentage point.}
\label{tab:evidtop10}
\end{table}

These inference-based results reinforce the central role of the energy sector in driving EUA prices, while also illustrating how broader financial signals, including equity performance and market volatility, mediate expectations around carbon demand.

\subsection{Sensitivity analysis}
We complement the inference results with a set of sensitivity measures that evaluate the strength and robustness of the learned relationships. These metrics focus on both structural and probabilistic aspects of the model and help identify which variables play the most influential roles in determining EUA futures prices.

We begin by examining arc strengths as depicted in Figure~\ref{fig:bayesnet}, where edge widths are proportional to the strength of each dependency. The strongest connections are observed between key financial market indicators; for example, from the S\&P 500 (SPX) to the STOXX Europe 600 (SXXP) and from the CAC 40 to the DAX Index, highlighting strong regional and transatlantic equity co-movement. Other prominent arcs include ECO to VIX and SPGTCED to ECO, pointing to tight interdependencies between clean energy and market volatility. In contrast, the arc strengths from MSCI Europe Energy (MXEU0EN) to coal (XA1) and from MXEU0EN to EUA futures (MO1) are relatively modest, suggesting that while energy sector dynamics are central to the structure, EUA pricing is influenced by a broader set of interacting signals rather than dominated by any single upstream relationship.

\begin{table}
\centering
\renewcommand{\arraystretch}{1.2}
\footnotesize
\scalebox{0.75}{
\begin{tabular}{lcc}
\toprule
\textbf{Variable} & \textbf{Mutual Info $\times$ 100 (Rank)} & \textbf{Sobol Index (Rank)} \\
\midrule
XA1 (Coal Futures)         & 0.947 (2)  & 0.470 (1) \\
MXEU0EN (MSCI Energy)      & 1.858 (1)  & 0.333 (2) \\
SXXP (STOXX Europe 600)    & 0.179 (6)  & 0.070 (3) \\
CAC Index                  & 0.334 (3)  & 0.067 (4) \\
DAX Index                  & 0.228 (7)  & 0.051 (5) \\
CO1 (Brent Oil)            & 0.003 (16) & 0.046 (6) \\
SPX (S\&P 500)             & 0.143 (10) & 0.016 (7) \\
EURRUB                     & 0.103 (9)  & 0.015 (8) \\
LP01TREU (High Yield)      & 0.037 (12) & 0.014 (9) \\
SPGTCED (S\&P Clean Energy)& 0.112 (8)  & 0.009 (10) \\
ECO Index                  & 0.055 (11) & 0.008 (11) \\
VIX                        & 0.037 (13) & 0.008 (12) \\
LBEATREU (Investment Grade)& 0.050 (10) & 0.003 (13) \\
EURGBP                     & 0.001 (14) & 0.001 (14) \\
EURUSD                     & 0.002 (15) & 0.000 (15) \\
EURCNY                     & 0.000 (17) & 0.000 (15) \\
EURCHF                     & 0.000 (17) & 0.000 (15) \\
XAU (Gold Spot)            & 0.007 (13) & 0.000 (15) \\
NG1COMB (Natural Gas)      & 0.000 (17) & 0.000 (15) \\
\bottomrule
\end{tabular}}
\caption{Mutual information (multiplied by 100) and Sobol indices for each variable with respect to EUA futures (MO1). Variables are ordered by decreasing Sobol index; ranks in each metric are shown in parentheses.}
\label{tab:sobol_mi}
\end{table}

We assess the marginal relevance of each variable using two complementary metrics: mutual information and the Sobol variance index. As shown in Table~\ref{tab:sobol_mi}, the coal futures (XA1) and MSCI Europe Energy Index (MXEU0EN) emerge as the most influential factors in explaining variability in EUA returns, consistently ranking highest under both criteria. These variables capture direct energy sector dynamics, confirming their dominant role in shaping carbon prices. Stock indices such as the STOXX Europe 600, CAC 40, and DAX also show high Sobol scores, suggesting that macroeconomic and equity market conditions substantially contribute to EUA uncertainty. In contrast, the mutual information metric, reflecting general dependence, assigns a higher rank to MSCI Energy but slightly downweights equity indices, indicating that their influence may be more specific to particular regimes. Currency pairs and volatility or safe-haven indicators (e.g., EURUSD, gold, VIX) show negligible sensitivity under both measures, highlighting their limited role in driving daily EUA price changes. These results reinforce the view that EUA pricing is primarily governed by energy fundamentals and market-wide financial signals.


We conclude the sensitivity analysis with tornado plots that visualize the most influential parameters affecting each outcome of the EUA node. The most sensitive parameters in all cases belong to the CPT of MO1 itself, consistent with the modest arc strength from its parent nodes. The most impactful configurations typically occur when both parents (XA1 and MXEU0EN) are fixed at the same level, highlighting the importance of aligned energy signals. For the high return case, additional sensitivity is observed when XA1 is set to “high”, reinforcing the role of coal in upward EUA movements. Compared to the neutral scenario, the high and low outcomes show more pronounced sensitivity patterns, suggesting that extreme EUA price shifts are more responsive to energy-related uncertainty. Full tornado plots for each outcome are reported in~\ref{appendix:tornado}.

\subsection{Temporal effects in the dynamic Bayesian network}

\begin{figure}
\centering
\includegraphics[width=.75\textwidth]{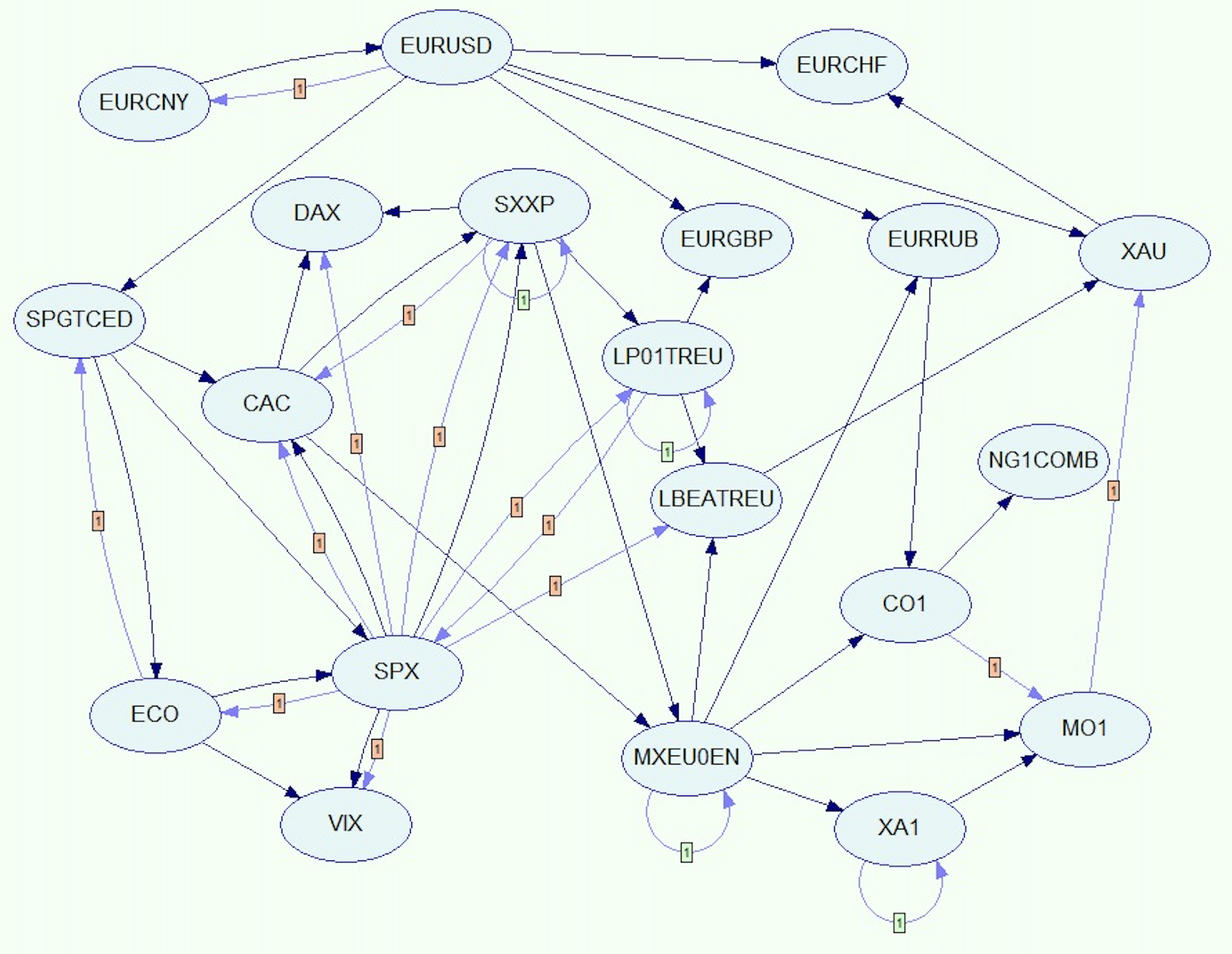}
\caption{Structure of the DBN. Arrows indicate dependencies between variable states at consecutive time steps ($T$ to $T+1$) and within the same time slice. The visualization follows the standard layout generated by GeNIe software.}
\label{fig:dbn}
\end{figure}

The learned DBN reveals how shocks to key variables influence market conditions on the following day. The estimated structure, shown in Figure~\ref{fig:dbn}, includes 17 inter-temporal edges from time $T$ to $T+1$, alongside several intra-slice dependencies. Four variables also exhibit self-loops, capturing significant autocorrelation across days.

The S\&P 500 index plays a central role in driving next-day outcomes, with outgoing edges to several U.S. and European market indicators, including the ECO and SPGTCED clean energy indices, the VIX, the STOXX Europe 600, and both investment- and high-yield bond indices. This highlights the predictive importance of the U.S. equity market for European financial variables at daily resolution.

Turning to the EUA futures node, only two meaningful inter-temporal links were detected. Brent oil futures influence next-day EUA probabilities, while EUA futures themselves transmit information to the gold spot price. Although the latter may be spurious, the former aligns with expectations: oil prices are closely tied to energy markets and investor sentiment, both of which affect EUA demand. Other potential predictors, including currency pairs and bond indices, show minimal or no influence on next-day EUA movements.

To assess the practical relevance of these temporal connections, we examined how shocks to selected variables affect the predicted distribution of EUA returns at both $T$ and $T+1$. Results for Brent oil and coal futures, the two most directly linked energy inputs, are shown in Table~\ref{tab:dbneffect}. Each row reports the conditional distribution of EUA returns (high, neutral, low) given a high or low return on the influencing variable.

\begin{table}
\centering
\footnotesize
\renewcommand{\arraystretch}{1.3}
\begin{tabular}{l|ccc|ccc}
\toprule
\multirow{2}{*}{\textbf{Evidence}} &
\multicolumn{3}{c|}{\textbf{EUA Prob. at $T$}} &
\multicolumn{3}{c}{\textbf{EUA Prob. at $T+1$}} \\
& High & Neutral & Low & High & Neutral & Low \\
\midrule
Oil Futures = High & 36\% & 33\% & 30\% & 30\% & 34\% & 35\% \\
Oil Futures = Low & 31\% & 33\% & 36\% & 35\% & 32\% & 31\% \\
Coal Futures = High & 42\% & 32\% & 24\% & 34\% & 33\% & 32\% \\
Coal Futures = Low & 24\% & 32\% & 43\% & 32\% & 33\% & 34\% \\
\bottomrule
\end{tabular}
\caption{Conditional distribution of EUA returns at $T$ and $T+1$ given high or low values in coal and oil futures at $T$.}
\label{tab:dbneffect}
\end{table}

Oil futures exert a small but measurable time-lagged influence: a high oil return today decreases the chance of a high EUA return tomorrow (30\%), while a drop in oil price increases it (35\%). This effect, though limited in magnitude, is consistent with market expectations that rising oil prices may dampen industrial activity and carbon demand. Coal futures, by contrast, affect EUA pricing predominantly within the same day. The time-lagged impact on $T+1$ is weaker and closer to the baseline distribution, suggesting that most of the coal-related signal is captured contemporaneously.

Across the broader network, other temporal effects were weak or negligible. Changes in equity indices, bond markets, and foreign exchange variables at $T$ did influence next-day values within their own clusters but showed little propagation to EUA futures. These results highlight the limited role of inter-temporal dynamics in daily EUA forecasting, most dependencies are best captured in contemporaneous relationships.

\section{Discussion}

\subsection{Carbon pricing at the intersection of energy, equity, and macro-financial dynamics}

Our results confirm that EUA pricing is influenced by interconnected factors across energy, equity, and macro-financial markets. The structure of the learned BN reflects known patterns: equity indices form two coherent regional clusters, with the S\&P 500 connected to both the STOXX 600 and clean energy indices. This is consistent with prior studies highlighting strong co-movements in investor sentiment and economic expectations across U.S. and European markets \citep{wang2020, hanif2021, salvagnin2024}.

A second major cluster centers on the MSCI Europe Energy Index, which functions as a bridge between fossil fuel markets and EUA prices. This index connects directly to coal and oil futures and influences bond markets, underscoring the energy sector's centrality in emissions trading. The observed links align with findings that energy sector performance is a key compliance signal under the ETS \citep{lovcha2021}, but our model offers a more comprehensive view of how energy equity interacts with commodity and financial variables.

Foreign exchange pairs form a separate, largely self-contained cluster. While the U.S. Dollar influences multiple currencies, its direct connection to EUA pricing appears limited. This suggests that foreign exchange effects on EUA prices are generally indirect, mediated through their impact on commodity or equity markets. Our results indicate that these macro-financial influences are present but relatively modest in comparison to more proximal drivers.

Bond markets also play a role, with the MSCI Europe Energy Index influencing investment-grade bond indices. Although this direction is not commonly highlighted in existing studies, it suggests that bond markets respond to energy sector dynamics more than previously recognized, particularly as carbon-intensive firms carry higher regulatory exposure.

\subsection{Consistent signals and new insights on EUA price formation}

Our analysis of probabilistic effects confirms the dominant role of energy markets in determining EUA price shifts. Among all variables, coal futures exert the strongest effect: a high coal price increases the probability of a high EUA return by nearly 20 percentage points relative to a low coal price. This supports the fuel-switching mechanism but adds nuance, indicating that rising coal prices, in a short-term context, are interpreted as tightening energy supply and increasing allowance demand.

The role of the MSCI Europe Energy Index is also novel in its structural centrality. While earlier studies noted its relevance \citep{salvagnin2024}, our network shows that it mediates effects between energy commodities, financial indices, and the EUA market. This suggests that equity-based performance of regulated firms acts as a meaningful proxy for expected compliance behavior under the ETS.

European stock indices, especially the STOXX 600, CAC 40, and DAX, also influence EUA returns, confirming macroeconomic expectations as key short-term drivers. Their inclusion supports the findings of \citet{tan2017} and \citet{wang2020}, which show that EUA prices rise alongside industrial activity and investor optimism.

Clean energy indices (ECO and SPGTCED) show weaker but positive effects. While these sectors are expected to reduce long-term EUA demand, our findings support the view that in the short run, rising clean energy equity may reflect broader investment sentiment rather than structural decarbonization. This is in line with \citet{hanif2021}, who identify these indices as short-term volatility contributors rather than consistent causal drivers.

Currencies and gold show negligible influence, reaffirming the idea that carbon markets are primarily shaped by energy and production-related variables. The limited role of foreign exchange contrasts with some earlier claims, but supports more recent studies that emphasize domestic energy signals and compliance behavior as the key forces behind EUA price formation.

\subsection{Temporal transmission and limitations of dynamic effects}

The DBN adds a temporal dimension, revealing that most effects influencing EUA futures occur contemporaneously. Very few inter-temporal effects are identified, with Brent oil futures being the only variable to systematically affect EUA prices on the following day. This modest lag likely reflects how oil shocks inform short-term energy pricing and industrial sentiment, as noted in \citet{zheng2021}.

Other temporal edges, such as the influence of the S\&P 500 on European markets have little predictive power for EUA pricing. Most other variables, including equity indices, bonds, and foreign exchange, do not transmit meaningful signals to EUA returns beyond the current day.

Autocorrelation patterns within variables, particularly coal and energy indices, are strong. However, their influence on EUA returns does not persist across time steps. This confirms expectations from high-frequency financial models: while variables may exhibit persistence, most relevant information is rapidly priced in, and past values have limited additional predictive value.


Taken together, the dynamic results underscore a key contribution of this study: while EUA prices are driven by a complex system of interdependent factors, the market adjusts rapidly, and most meaningful information is embedded in the contemporaneous structure. This suggests that forecasting approaches should prioritize structural inference over lag-based signal extraction.

\section{Conclusion}

This study has examined the short-term drivers of EUA futures prices using both static and dynamic BNs, providing a probabilistic perspective on how energy, financial, and macroeconomic variables interact in shaping the carbon market. Our findings confirm the central role of the energy sector: coal and oil prices, along with the equity performance of energy-intensive firms, emerge as the most influential drivers of EUA returns. Broader macro-financial conditions, as captured by equity indices and bond markets, also shape EUA pricing, primarily through their effects on energy demand and investor sentiment.

While the static network captures a rich structure of contemporaneous dependencies, the dynamic extension reveals that inter-temporal effects are limited. Apart from modest next-day influences, notably from oil futures, most information appears to be rapidly incorporated into EUA prices, underscoring the responsiveness of carbon markets to new data.

By integrating a diverse set of variables into a unified, interpretable framework, our approach contributes to the literature on carbon pricing by identifying and quantifying key interdependencies. The resulting insights can support ETS-regulated firms in designing hedging strategies, assist investors in anticipating EUA price movements, and inform policymakers about the broader financial and energy dynamics affecting allowance demand.

Several avenues remain for further research. The current DBN framework is limited to first-order temporal effects; higher-order or time-varying models could uncover additional dynamics. The discretization scheme, while enhancing interpretability, may be refined through alternative binning strategies or extended to continuous-variable formulations. Incorporating regulatory announcements, firm-level compliance data, or auction outcomes could also provide greater granularity in modeling market reactions.

Future work might also explore volatility dynamics using multivariate GARCH models \citep{bauwens2006multivariate} or assess the role of extreme events through graphical approaches to extreme value theory \citep{engelke2020graphical}. These extensions would help capture asymmetries and tail risks that are increasingly relevant in turbulent energy and carbon markets.

\bibliographystyle{elsarticle-harv} 
 \bibliography{bib}

\begin{thebibliography}{46}
\expandafter\ifx\csname natexlab\endcsname\relax\def\natexlab#1{#1}\fi
\providecommand{\url}[1]{\texttt{#1}}
\providecommand{\href}[2]{#2}
\providecommand{\path}[1]{#1}
\providecommand{\DOIprefix}{doi:}
\providecommand{\ArXivprefix}{arXiv:}
\providecommand{\URLprefix}{URL: }
\providecommand{\Pubmedprefix}{pmid:}
\providecommand{\doi}[1]{\href{http://dx.doi.org/#1}{\path{#1}}}
\providecommand{\Pubmed}[1]{\href{pmid:#1}{\path{#1}}}
\providecommand{\bibinfo}[2]{#2}
\ifx\xfnm\relax \def\xfnm[#1]{\unskip,\space#1}\fi
\bibitem[{Ankan and Textor(2024)}]{ankan2024}
\bibinfo{author}{Ankan, A.}, \bibinfo{author}{Textor, J.}, \bibinfo{year}{2024}.
\newblock \bibinfo{title}{{Pgmpy: A Python toolkit for Bayesian networks}}.
\newblock \bibinfo{journal}{Journal of Machine Learning Research} \bibinfo{volume}{25}, \bibinfo{pages}{1--8}.
\bibitem[{Ballester-Ripoll and Leonelli(2022)}]{ballester2022computing}
\bibinfo{author}{Ballester-Ripoll, R.}, \bibinfo{author}{Leonelli, M.}, \bibinfo{year}{2022}.
\newblock \bibinfo{title}{Computing {S}obol indices in probabilistic graphical models}.
\newblock \bibinfo{journal}{Reliability Engineering \& System Safety} \bibinfo{volume}{225}, \bibinfo{pages}{108573}.
\bibitem[{Ballester-Ripoll and Leonelli(2023)}]{ballester2023yodo}
\bibinfo{author}{Ballester-Ripoll, R.}, \bibinfo{author}{Leonelli, M.}, \bibinfo{year}{2023}.
\newblock \bibinfo{title}{{The YODO algorithm: An efficient computational framework for sensitivity analysis in Bayesian networks}}.
\newblock \bibinfo{journal}{International Journal of Approximate Reasoning} \bibinfo{volume}{159}, \bibinfo{pages}{108929}.
\bibitem[{Barons et~al.(2022)Barons, Mascaro and Hanea}]{barons2022balancing}
\bibinfo{author}{Barons, M.J.}, \bibinfo{author}{Mascaro, S.}, \bibinfo{author}{Hanea, A.M.}, \bibinfo{year}{2022}.
\newblock \bibinfo{title}{Balancing the elicitation burden and the richness of expert input when quantifying discrete {B}ayesian networks}.
\newblock \bibinfo{journal}{Risk Analysis} \bibinfo{volume}{42}, \bibinfo{pages}{1196--1234}.
\bibitem[{Bauwens et~al.(2006)Bauwens, Laurent and Rombouts}]{bauwens2006multivariate}
\bibinfo{author}{Bauwens, L.}, \bibinfo{author}{Laurent, S.}, \bibinfo{author}{Rombouts, J.V.}, \bibinfo{year}{2006}.
\newblock \bibinfo{title}{{Multivariate GARCH models: A survey}}.
\newblock \bibinfo{journal}{Journal of Applied Econometrics} \bibinfo{volume}{21}, \bibinfo{pages}{79--109}.
\bibitem[{Beuzen et~al.(2018)Beuzen, Marshall and Splinter}]{beuzen2018}
\bibinfo{author}{Beuzen, T.}, \bibinfo{author}{Marshall, L.}, \bibinfo{author}{Splinter, K.D.}, \bibinfo{year}{2018}.
\newblock \bibinfo{title}{A comparison of methods for discretizing continuous variables in {B}ayesian networks}.
\newblock \bibinfo{journal}{Environmental Modelling \& Software} \bibinfo{volume}{108}, \bibinfo{pages}{61--66}.
\bibitem[{Borghesi et~al.(2023)Borghesi, Pahle, Perino, Quemin and Willner}]{borghesi2023}
\bibinfo{author}{Borghesi, S.}, \bibinfo{author}{Pahle, M.}, \bibinfo{author}{Perino, G.}, \bibinfo{author}{Quemin, S.}, \bibinfo{author}{Willner, M.}, \bibinfo{year}{2023}.
\newblock \bibinfo{title}{The market stability reserve in the {EU} emissions trading system: A critical review}.
\newblock \bibinfo{journal}{Annual Review of Resource Economics} \bibinfo{volume}{15}, \bibinfo{pages}{131--152}.
\bibitem[{Borunda et~al.(2016)Borunda, Jaramillo, Reyes and Ibarg{\"u}engoytia}]{borunda2016bayesian}
\bibinfo{author}{Borunda, M.}, \bibinfo{author}{Jaramillo, O.}, \bibinfo{author}{Reyes, A.}, \bibinfo{author}{Ibarg{\"u}engoytia, P.H.}, \bibinfo{year}{2016}.
\newblock \bibinfo{title}{Bayesian networks in renewable energy systems: A bibliographical survey}.
\newblock \bibinfo{journal}{Renewable and Sustainable Energy Reviews} \bibinfo{volume}{62}, \bibinfo{pages}{32--45}.
\bibitem[{Chickering(2002)}]{chickering2002learning}
\bibinfo{author}{Chickering, D.M.}, \bibinfo{year}{2002}.
\newblock \bibinfo{title}{Learning equivalence classes of {B}ayesian-network structures}.
\newblock \bibinfo{journal}{Journal of Machine Learning Research} \bibinfo{volume}{2}, \bibinfo{pages}{445--498}.
\bibitem[{Engelke and Hitz(2020)}]{engelke2020graphical}
\bibinfo{author}{Engelke, S.}, \bibinfo{author}{Hitz, A.S.}, \bibinfo{year}{2020}.
\newblock \bibinfo{title}{Graphical models for extremes}.
\newblock \bibinfo{journal}{Journal of the Royal Statistical Society Series B} \bibinfo{volume}{82}, \bibinfo{pages}{871--932}.
\bibitem[{Epskamp et~al.(2012)Epskamp, Cramer, Waldorp, Schmittmann and Borsboom}]{epskamp2012qgraph}
\bibinfo{author}{Epskamp, S.}, \bibinfo{author}{Cramer, A.O.}, \bibinfo{author}{Waldorp, L.J.}, \bibinfo{author}{Schmittmann, V.D.}, \bibinfo{author}{Borsboom, D.}, \bibinfo{year}{2012}.
\newblock \bibinfo{title}{qgraph: Network visualizations of relationships in psychometric data}.
\newblock \bibinfo{journal}{Journal of Statistical Software} \bibinfo{volume}{48}, \bibinfo{pages}{1--18}.
\bibitem[{{European Commission}()}]{ec_ets}
\bibinfo{author}{{European Commission}}, .
\newblock \bibinfo{title}{Eu emissions trading system (eu ets)}.
\newblock \URLprefix \url{https://climate.ec.europa.eu/eu-action/eu-emissions-trading-system-eu-ets_en}. \bibinfo{note}{accessed: 2025-04-24}.
\bibitem[{{European Environment Agency}()}]{eea}
\bibinfo{author}{{European Environment Agency}}, .
\newblock \bibinfo{title}{Eu emissions trading system (eu ets) data viewer: Explore emissions and allowances}.
\newblock \URLprefix \url{https://www.eea.europa.eu/en/analysis/maps-and-charts/emissions-trading-viewer-1-dashboards}. \bibinfo{note}{accessed: 2025-04-24}.
\bibitem[{{European Securities and Markets Authority}(2024)}]{esma2024}
\bibinfo{author}{{European Securities and Markets Authority}}, \bibinfo{year}{2024}.
\newblock \bibinfo{title}{Carbon markets report 2024}.
\newblock \URLprefix \url{https://www.esma.europa.eu/sites/default/files/2024-10/ESMA50-43599798-10379_Carbon_markets_report_2024.pdf}.
\bibitem[{Hailemariam et~al.(2022)Hailemariam, Ivanovski and Dzhumashev}]{hailemariam2022}
\bibinfo{author}{Hailemariam, A.}, \bibinfo{author}{Ivanovski, K.}, \bibinfo{author}{Dzhumashev, R.}, \bibinfo{year}{2022}.
\newblock \bibinfo{title}{Does {R\&D} investment in renewable energy technologies reduce greenhouse gas emissions?}
\newblock \bibinfo{journal}{Applied Energy} \bibinfo{volume}{327}, \bibinfo{pages}{120056}.
\bibitem[{Hanif et~al.(2021)Hanif, Hernandez, Mensi, Kang, Uddin and Yoon}]{hanif2021}
\bibinfo{author}{Hanif, W.}, \bibinfo{author}{Hernandez, J.A.}, \bibinfo{author}{Mensi, W.}, \bibinfo{author}{Kang, S.H.}, \bibinfo{author}{Uddin, G.S.}, \bibinfo{author}{Yoon, S.M.}, \bibinfo{year}{2021}.
\newblock \bibinfo{title}{Nonlinear dependence and connectedness between clean/renewable energy sector equity and {E}uropean emission allowance prices}.
\newblock \bibinfo{journal}{Energy Economics} \bibinfo{volume}{101}, \bibinfo{pages}{105409}.
\bibitem[{Heckerman et~al.(1995)Heckerman, Geiger and Chickering}]{heckerman1995learning}
\bibinfo{author}{Heckerman, D.}, \bibinfo{author}{Geiger, D.}, \bibinfo{author}{Chickering, D.M.}, \bibinfo{year}{1995}.
\newblock \bibinfo{title}{Learning {B}ayesian networks: The combination of knowledge and statistical data}.
\newblock \bibinfo{journal}{Machine Learning} \bibinfo{volume}{20}, \bibinfo{pages}{197--243}.
\bibitem[{{ICIS}(2020)}]{icis2020}
\bibinfo{author}{{ICIS}}, \bibinfo{year}{2020}.
\newblock \bibinfo{title}{The eua market: Understanding price formation and risks in the eu ets}.
\newblock \URLprefix \url{https://sparkchange.io/wp-content/uploads/2023/03/ICIS-EU-ETS-Report_2020-11-26_The-EUA-Market148054.pdf}.
\bibitem[{Kaikkonen et~al.(2020)Kaikkonen, Parviainen, Rahikainen, Uusitalo and Lehikoinen}]{kaikkonen2020bayesian}
\bibinfo{author}{Kaikkonen, L.}, \bibinfo{author}{Parviainen, T.}, \bibinfo{author}{Rahikainen, M.}, \bibinfo{author}{Uusitalo, L.}, \bibinfo{author}{Lehikoinen, A.}, \bibinfo{year}{2020}.
\newblock \bibinfo{title}{Bayesian networks in environmental risk assessment: A review}.
\newblock \bibinfo{journal}{Integrated Environmental Assessment and Management} \bibinfo{volume}{17}, \bibinfo{pages}{62--78}.
\bibitem[{Kitson et~al.(2023)Kitson, Constantinou, Guo, Liu and Chobtham}]{kitson2023}
\bibinfo{author}{Kitson, N.K.}, \bibinfo{author}{Constantinou, A.C.}, \bibinfo{author}{Guo, Z.}, \bibinfo{author}{Liu, Y.}, \bibinfo{author}{Chobtham, K.}, \bibinfo{year}{2023}.
\newblock \bibinfo{title}{A survey of {Bayesian} network structure learning}.
\newblock \bibinfo{journal}{Artificial Intelligence Review} \bibinfo{volume}{56}, \bibinfo{pages}{8721--8814}.
\bibitem[{Koch et~al.(2014)Koch, Fuss, Grosjean and Edenhofer}]{koch2014}
\bibinfo{author}{Koch, N.}, \bibinfo{author}{Fuss, S.}, \bibinfo{author}{Grosjean, G.}, \bibinfo{author}{Edenhofer, O.}, \bibinfo{year}{2014}.
\newblock \bibinfo{title}{{Causes of the EU ETS price drop: Recession, CDM, renewable policies or a bit of everything?—New evidence}}.
\newblock \bibinfo{journal}{Energy Policy} \bibinfo{volume}{73}, \bibinfo{pages}{676--685}.
\bibitem[{Koller and Friedman(2009)}]{koller2009}
\bibinfo{author}{Koller, D.}, \bibinfo{author}{Friedman, N.}, \bibinfo{year}{2009}.
\newblock \bibinfo{title}{Probabilistic Graphical Models: Principles and Techniques}.
\newblock \bibinfo{publisher}{MIT Press}.
\bibitem[{Kwisthout(2011)}]{kwisthout2011most}
\bibinfo{author}{Kwisthout, J.}, \bibinfo{year}{2011}.
\newblock \bibinfo{title}{Most probable explanations in {B}ayesian networks: Complexity and tractability}.
\newblock \bibinfo{journal}{International Journal of Approximate Reasoning} \bibinfo{volume}{52}, \bibinfo{pages}{1452--1469}.
\bibitem[{Leitao et~al.(2021)Leitao, Ferreira and Santibanez-Gonzalez}]{leitao2021}
\bibinfo{author}{Leitao, J.}, \bibinfo{author}{Ferreira, J.}, \bibinfo{author}{Santibanez-Gonzalez, E.}, \bibinfo{year}{2021}.
\newblock \bibinfo{title}{Green bonds, sustainable development and environmental policy in the {European Union} carbon market}.
\newblock \bibinfo{journal}{Business Strategy and the Environment} \bibinfo{volume}{30}, \bibinfo{pages}{2077--2090}.
\bibitem[{Leonelli(2025)}]{leonelli2025bnrep}
\bibinfo{author}{Leonelli, M.}, \bibinfo{year}{2025}.
\newblock \bibinfo{title}{bnrep: A repository of {B}ayesian networks from the academic literature}.
\newblock \bibinfo{journal}{Neurocomputing} \bibinfo{volume}{624}, \bibinfo{pages}{129502}.
\bibitem[{Leonelli et~al.(2023)Leonelli, Ramanathan and Wilkerson}]{leonelli2023sensitivity}
\bibinfo{author}{Leonelli, M.}, \bibinfo{author}{Ramanathan, R.}, \bibinfo{author}{Wilkerson, R.L.}, \bibinfo{year}{2023}.
\newblock \bibinfo{title}{Sensitivity and robustness analysis in {B}ayesian networks with the bnmonitor {R} package}.
\newblock \bibinfo{journal}{Knowledge-Based Systems} \bibinfo{volume}{278}, \bibinfo{pages}{110882}.
\bibitem[{Leonelli and Smith(2025)}]{leonelli2025diameter}
\bibinfo{author}{Leonelli, M.}, \bibinfo{author}{Smith, J.}, \bibinfo{year}{2025}.
\newblock \bibinfo{title}{The diameter of a stochastic matrix: A new measure for sensitivity analysis in {B}ayesian networks}.
\newblock \bibinfo{journal}{International Journal of Approximate Reasoning (to appear)} .
\bibitem[{Lovcha et~al.(2021)Lovcha, Perez-Laborda and Sikora}]{lovcha2021}
\bibinfo{author}{Lovcha, Y.}, \bibinfo{author}{Perez-Laborda, A.}, \bibinfo{author}{Sikora, I.}, \bibinfo{year}{2021}.
\newblock \bibinfo{title}{The determinants of {CO2} prices in the {EU} emission trading system}.
\newblock \bibinfo{journal}{Applied Energy} \bibinfo{volume}{305}, \bibinfo{pages}{117903}.
\bibitem[{Machado et~al.(2023)Machado, de~Oliveira~Ribeiro and do~Nascimento}]{machado2023risk}
\bibinfo{author}{Machado, P.G.}, \bibinfo{author}{de~Oliveira~Ribeiro, C.}, \bibinfo{author}{do~Nascimento, C.A.O.}, \bibinfo{year}{2023}.
\newblock \bibinfo{title}{Risk analysis in energy projects using {B}ayesian networks: A systematic review}.
\newblock \bibinfo{journal}{Energy Strategy Reviews} \bibinfo{volume}{47}, \bibinfo{pages}{101097}.
\bibitem[{McNeil and Frey(2000)}]{mcneil2000estimation}
\bibinfo{author}{McNeil, A.J.}, \bibinfo{author}{Frey, R.}, \bibinfo{year}{2000}.
\newblock \bibinfo{title}{Estimation of tail-related risk measures for heteroscedastic financial time series: an extreme value approach}.
\newblock \bibinfo{journal}{Journal of Empirical Finance} \bibinfo{volume}{7}, \bibinfo{pages}{271--300}.
\bibitem[{Moe et~al.(2020)Moe, Carriger and Glendell}]{moe2020increased}
\bibinfo{author}{Moe, S.J.}, \bibinfo{author}{Carriger, J.F.}, \bibinfo{author}{Glendell, M.}, \bibinfo{year}{2020}.
\newblock \bibinfo{title}{Increased use of {B}ayesian network models has improved environmental risk assessments}.
\newblock \bibinfo{journal}{Integrated Environmental Assessment and Management} \bibinfo{volume}{17}, \bibinfo{pages}{53--61}.
\bibitem[{Murphy(2002)}]{murphy2002}
\bibinfo{author}{Murphy, K.P.}, \bibinfo{year}{2002}.
\newblock \bibinfo{title}{Dynamic Bayesian Networks: Representation, Inference and Learning}.
\newblock Ph.D. thesis. University of California, Berkeley.
\bibitem[{Nojavan et~al.(2017)Nojavan, Qian and Stow}]{nojavan2017}
\bibinfo{author}{Nojavan, F.A.}, \bibinfo{author}{Qian, S.S.}, \bibinfo{author}{Stow, C.A.}, \bibinfo{year}{2017}.
\newblock \bibinfo{title}{Comparative analysis of discretization methods in {B}ayesian networks}.
\newblock \bibinfo{journal}{Environmental Modelling \& Software} \bibinfo{volume}{87}, \bibinfo{pages}{64--71}.
\bibitem[{Nyberg et~al.(2022)Nyberg, Nicholson, Korb, Wybrow, Zukerman, Mascaro, Thakur, Oshni~Alvandi, Riley, Pearson et~al.}]{nyberg2022bard}
\bibinfo{author}{Nyberg, E.P.}, \bibinfo{author}{Nicholson, A.E.}, \bibinfo{author}{Korb, K.B.}, \bibinfo{author}{Wybrow, M.}, \bibinfo{author}{Zukerman, I.}, \bibinfo{author}{Mascaro, S.}, \bibinfo{author}{Thakur, S.}, \bibinfo{author}{Oshni~Alvandi, A.}, \bibinfo{author}{Riley, J.}, \bibinfo{author}{Pearson, R.}, et~al., \bibinfo{year}{2022}.
\newblock \bibinfo{title}{{BARD: A structured technique for group elicitation of Bayesian networks to support analytic reasoning}}.
\newblock \bibinfo{journal}{Risk Analysis} \bibinfo{volume}{42}, \bibinfo{pages}{1155--1178}.
\bibitem[{Pearl(1988)}]{pearl2014}
\bibinfo{author}{Pearl, J.}, \bibinfo{year}{1988}.
\newblock \bibinfo{title}{Probabilistic Reasoning in Intelligent Systems: Networks of Plausible Inference}.
\newblock \bibinfo{publisher}{Morgan Kaufmann}.
\bibitem[{Pe{\~n}a and Rodr{\'\i}guez(2019)}]{pena2022}
\bibinfo{author}{Pe{\~n}a, J.I.}, \bibinfo{author}{Rodr{\'\i}guez, R.}, \bibinfo{year}{2019}.
\newblock \bibinfo{title}{{Are EU's Climate and Energy Package 20-20-20 targets achievable and compatible? Evidence from the impact of renewables on electricity prices}}.
\newblock \bibinfo{journal}{Energy} \bibinfo{volume}{183}, \bibinfo{pages}{477--486}.
\bibitem[{Rudin(2019)}]{rudin2019stop}
\bibinfo{author}{Rudin, C.}, \bibinfo{year}{2019}.
\newblock \bibinfo{title}{Stop explaining black box machine learning models for high stakes decisions and use interpretable models instead}.
\newblock \bibinfo{journal}{Nature Machine Intelligence} \bibinfo{volume}{1}, \bibinfo{pages}{206--215}.
\bibitem[{Ruppert and Matteson(2015)}]{ruppert2015}
\bibinfo{author}{Ruppert, D.}, \bibinfo{author}{Matteson, D.}, \bibinfo{year}{2015}.
\newblock \bibinfo{title}{Statistics and Data Analysis for Financial Engineering}.
\newblock \bibinfo{edition}{2} ed., \bibinfo{publisher}{Springer}.
\bibitem[{Salvagnin et~al.(2024)Salvagnin, Glielmo, De~Giuli and Mira}]{salvagnin2024}
\bibinfo{author}{Salvagnin, C.}, \bibinfo{author}{Glielmo, A.}, \bibinfo{author}{De~Giuli, M.E.}, \bibinfo{author}{Mira, A.}, \bibinfo{year}{2024}.
\newblock \bibinfo{title}{Investigating the price determinants of the {European Emission Trading System:} a non-parametric approach}.
\newblock \bibinfo{journal}{Quantitative Finance} \bibinfo{volume}{24}, \bibinfo{pages}{1529--1544}.
\bibitem[{Scutari et~al.(2019)Scutari, Graafland and Gutiérrez}]{scutari2019}
\bibinfo{author}{Scutari, M.}, \bibinfo{author}{Graafland, C.E.}, \bibinfo{author}{Gutiérrez, J.M.}, \bibinfo{year}{2019}.
\newblock \bibinfo{title}{Who learns better {B}ayesian network structures: Accuracy and speed of structure learning algorithms}.
\newblock \bibinfo{journal}{International Journal of Approximate Reasoning} \bibinfo{volume}{115}, \bibinfo{pages}{235--253}.
\bibitem[{Scutari and Nagarajan(2013)}]{scutari2013}
\bibinfo{author}{Scutari, M.}, \bibinfo{author}{Nagarajan, R.}, \bibinfo{year}{2013}.
\newblock \bibinfo{title}{Identifying significant edges in graphical models of molecular networks}.
\newblock \bibinfo{journal}{Artificial Intelligence in Medicine} \bibinfo{volume}{57}, \bibinfo{pages}{207--217}.
\bibitem[{Sheppard(2023)}]{sheppard2023arch}
\bibinfo{author}{Sheppard, K.}, \bibinfo{year}{2023}.
\newblock \bibinfo{title}{arch: Autoregressive Conditional Heteroskedasticity Models}.
\newblock \URLprefix \url{https://arch.readthedocs.io/}. \bibinfo{note}{python library version 6.4}.
\bibitem[{Tan and Wang(2017)}]{tan2017}
\bibinfo{author}{Tan, X.P.}, \bibinfo{author}{Wang, X.Y.}, \bibinfo{year}{2017}.
\newblock \bibinfo{title}{Dependence changes between the carbon price and its fundamentals: A quantile regression approach}.
\newblock \bibinfo{journal}{Applied Energy} \bibinfo{volume}{190}, \bibinfo{pages}{306--325}.
\bibitem[{Terranova et~al.(2024)Terranova, Cozzarini, Reissl and Tavoni}]{terranova2024}
\bibinfo{author}{Terranova, R.}, \bibinfo{author}{Cozzarini, C.}, \bibinfo{author}{Reissl, S.}, \bibinfo{author}{Tavoni, M.}, \bibinfo{year}{2024}.
\newblock \bibinfo{title}{Detecting speculation in the market for {EU} emission allowances}.
\newblock \bibinfo{note}{Available at SSRN 5030260}.
\bibitem[{Wang and Zhao(2020)}]{wang2020}
\bibinfo{author}{Wang, Z.}, \bibinfo{author}{Zhao, L.}, \bibinfo{year}{2020}.
\newblock \bibinfo{title}{The impact of the global stock and energy market on {EU ETS:} a structural equation modelling approach}.
\newblock \bibinfo{journal}{Journal of Cleaner Production} \bibinfo{volume}{289}, \bibinfo{pages}{125140}.
\bibitem[{Zheng et~al.(2021)Zheng, Yin, Zhou, Liu and Wen}]{zheng2021}
\bibinfo{author}{Zheng, Y.}, \bibinfo{author}{Yin, H.}, \bibinfo{author}{Zhou, M.}, \bibinfo{author}{Liu, W.}, \bibinfo{author}{Wen, F.}, \bibinfo{year}{2021}.
\newblock \bibinfo{title}{Impacts of oil shocks on the {EU} carbon emissions allowances under different market conditions}.
\newblock \bibinfo{journal}{Energy Economics} \bibinfo{volume}{104}, \bibinfo{pages}{105683}.

\end{thebibliography}






\appendix

\newpage
\section{Key terms used in the literature review}

\begin{table}[h]
    \centering
\renewcommand{\arraystretch}{1.4}
    \scalebox{0.66}{
\begin{tabular}{p{5cm}p{3cm}p{8cm}}
\toprule
\textbf{Term} & \textbf{Abbreviation} & \textbf{Explanation} \\
\midrule
Emissions Trading System & ETS & The European Union’s cap-and-trade system for regulating greenhouse gas emissions through market-based mechanisms. \\
European Union Allowance & EUA & A tradeable unit giving the holder the right to emit one tonne of carbon dioxide or its equivalent. \\
Fit for 55 & FF55 & A package of EU reforms aiming to reduce greenhouse gas emissions by 55\% by 2030 compared to 1990 levels. \\
European Energy Exchange & EEX & A German-based exchange platform for electricity and related energy products. \\
Futures contract & – & A financial derivative that obligates the exchange of a specified asset at a future date and predetermined price. \\
Market Stability Reserve & MSR & A mechanism that adjusts the volume of carbon allowances auctioned based on surplus in circulation. \\
Total Number of Allowances in Circulation & TNAC & A measure of all carbon allowances currently issued and available in the market. \\
Linear Reduction Factor & LRF & The annual percentage by which the cap on allowances is reduced under EU climate targets. \\
Fuel switching & – & The process of changing the base fuel used in energy production in response to cost or regulatory pressures. \\
Independent Commodity Intelligence Services & ICIS & A market analysis and reporting firm focusing on global commodity and energy markets. \\
European Securities and Markets Authority & ESMA & The EU authority responsible for regulating and supervising financial markets, including EUA trading. \\
Intercontinental Exchange & ICE & A global trading platform where most secondary EUA futures transactions take place. \\
\bottomrule
\end{tabular}
}
\caption{Glossary of terms used in the literature review.}
\label{tab:termglossary}
\end{table}

\newpage
\section{Summary of reviewed works}
\begin{table}[h]
    \centering
\renewcommand{\arraystretch}{1.4}
    \scalebox{0.66}{
\begin{tabular}{p{3.5cm}p{3.5cm}p{3.6cm}p{7.4cm}}
\toprule
\textbf{Work} & \textbf{Models} & \textbf{Features} & \textbf{Comments} \\
\midrule
Borghesi et al. (2023) & Conceptual MSR design analysis & – & Reviews past and future MSR trends. Price volatility linked to strategic banking and reduced supply. MSR is unresponsive to long-term demand shocks. \\
Hanif et al. (2021) & Spillover index, Copulas & 6 clean energy indices & EUA prices are net recipients of short-term spillovers from clean energy stocks, especially during market stress and regulatory events. \\
Leitao et al. (2021) & Markov-switching model & Green and conventional bond indices, commodity index & Finds positive effects of Green Bonds on EUA prices across regimes. Conventional bonds negatively affect EUA prices. \\
Lovcha et al. (2021) & SVAR, frequency analysis & Commodities, electricity, stock index, renewables share & Coal and gas prices matter due to fuel switching. EUA is input to electricity prices. Short-term speculation significant, long-term fundamentals dominate. \\
Peña \& Rodriguez (2022) & Panel regression, ARMA, Monte Carlo & Wholesale electricity prices & More renewables lower wholesale prices, which may increase electricity consumption and emissions. \\
Salvagnin et al. (2024) & Info imbalance, Gaussian processes & Macroeconomic and energy indicators & EUA demand linked to industrial activity and economic signals. Drivers have shifted post-COVID and amid regulatory reforms. \\
Tan \& Wang (2017) & Quantile regression & Energy commodities, foreign exchange, macro indicators & Inverse price relation via fuel switching. Risk influences energy markets and EUA demand. High-yield bonds reflect risk stress. \\
Terranova et al. (2024) & Bubble detection tests & EUA futures prices & Identifies seven short-term speculative bubbles linked to regulatory announcements. \\
Wang \& Zhao (2020) & Bayesian network, SEM & Energy and macro factors & Clean energy affects the energy sector and indirectly the EUA price via financial market interactions. \\
Zheng et al. (2021) & Quantile regression & Oil, gas, foreign exchange, VIX & Oil price volatility lowers EUA prices. High risk leads to higher yields, reduced production, and falling EUA demand. \\
\bottomrule
    \end{tabular}}
    \caption{Summary of key empirical works discussed in the literature review.}
\label{tab:litrevworks}
    \end{table}

\newpage

\section{Exploratory data analysis}
\label{appendix:eda}

The results of the exploratory data analysis confirm that the distribution of returns for all variables deviates substantially from a Gaussian shape and exhibits heavy tails. Descriptive statistics for all series are provided in Table~\ref{tab:descstats}. Non-normality is further supported by the Shapiro–Wilk test, which yields significant test statistics for all variables, rejecting the null hypothesis of normality in every case.

\begin{table}[h]
\centering
\renewcommand{\arraystretch}{1.4}
\footnotesize
\scalebox{0.88}{
\begin{tabular}{lcccccc}
\toprule
\textbf{Variable} & \textbf{Mean} & \textbf{Std. Dev.} & \textbf{Min} & \textbf{Max} & \textbf{Skewness} & \textbf{Exc. Kurtosis} \\
\midrule
MO1 Comdty & 0.0008 & 0.0312 & -0.4324 & 0.2392 & -0.98 & 13.40 \\
CO1 Comdty & -0.0001 & 0.0235 & -0.2798 & 0.1908 & -0.95 & 14.21 \\
XA1 Comdty & 0.0001 & 0.0251 & -0.5369 & 0.3262 & -2.83 & 95.62 \\
NG1 COMB Comdty & 0.0000 & 0.0370 & -0.3005 & 0.3817 & 0.19 & 4.76 \\
SPX Index & 0.0005 & 0.0105 & -0.1277 & 0.0897 & -0.83 & 14.05 \\
SXXP Index & 0.0002 & 0.0099 & -0.1219 & 0.0807 & -1.05 & 9.96 \\
MXEU0EN Index & 0.0000 & 0.0161 & -0.1994 & 0.1771 & -0.61 & 15.90 \\
SPGTCED Index & 0.0001 & 0.0145 & -0.1250 & 0.1103 & -0.35 & 4.76 \\
CAC Index & 0.0002 & 0.0115 & -0.1310 & 0.0806 & -0.79 & 7.76 \\
DAX Index & 0.0003 & 0.0116 & -0.1305 & 0.1041 & -0.59 & 7.45 \\
VIX Index & 0.0000 & 0.0775 & -0.3307 & 0.7682 & 1.30 & 4.80 \\
EURUSD Curncy & -0.0001 & 0.0048 & -0.0238 & 0.0302 & 0.07 & -0.87 \\
EURGBP Curncy & 0.0000 & 0.0047 & -0.0198 & 0.0601 & 0.87 & 7.40 \\
EURCHF Curncy & -0.0001 & 0.0049 & -0.2079 & 0.0292 & -23.99 & 1016.55 \\
EURCNY Curncy & -0.0000 & 0.0045 & -0.0213 & 0.0288 & 0.20 & -0.14 \\
EURRUB Curncy & 0.0003 & 0.0146 & -0.1257 & 0.2268 & 1.30 & 28.98 \\
ECO Index & -0.0000 & 0.0220 & -0.1624 & 0.1340 & -0.22 & 0.81 \\
LP01TREU Index & 0.0002 & 0.0025 & -0.0384 & 0.0197 & -3.63 & 49.76 \\
XAU Curncy & 0.0001 & 0.0092 & -0.0951 & 0.0497 & -0.58 & 3.46 \\
LBEATREU Index & 0.0000 & 0.0025 & -0.0129 & 0.0175 & 0.06 & 1.89 \\
\bottomrule
\end{tabular}
}
\caption{Descriptive statistics of daily log returns for all variables in the analysis.}
\label{tab:descstats}
\end{table}

Stationarity of the log-returns was assessed using both the Augmented Dickey–Fuller (ADF) and Kwiatkowski–Phillips–Schmidt–Shin (KPSS) tests. The ADF results reject the null hypothesis of a unit root for all variables, while the KPSS test confirms stationarity in all cases except for the gold spot price, where the p-value falls just below the standard 0.05 threshold. Given the mild rejection and the consistency of findings across both tests, we treat all series as stationary for the purposes of modeling.

\begin{figure}
\centering
\includegraphics[width=1\textwidth]{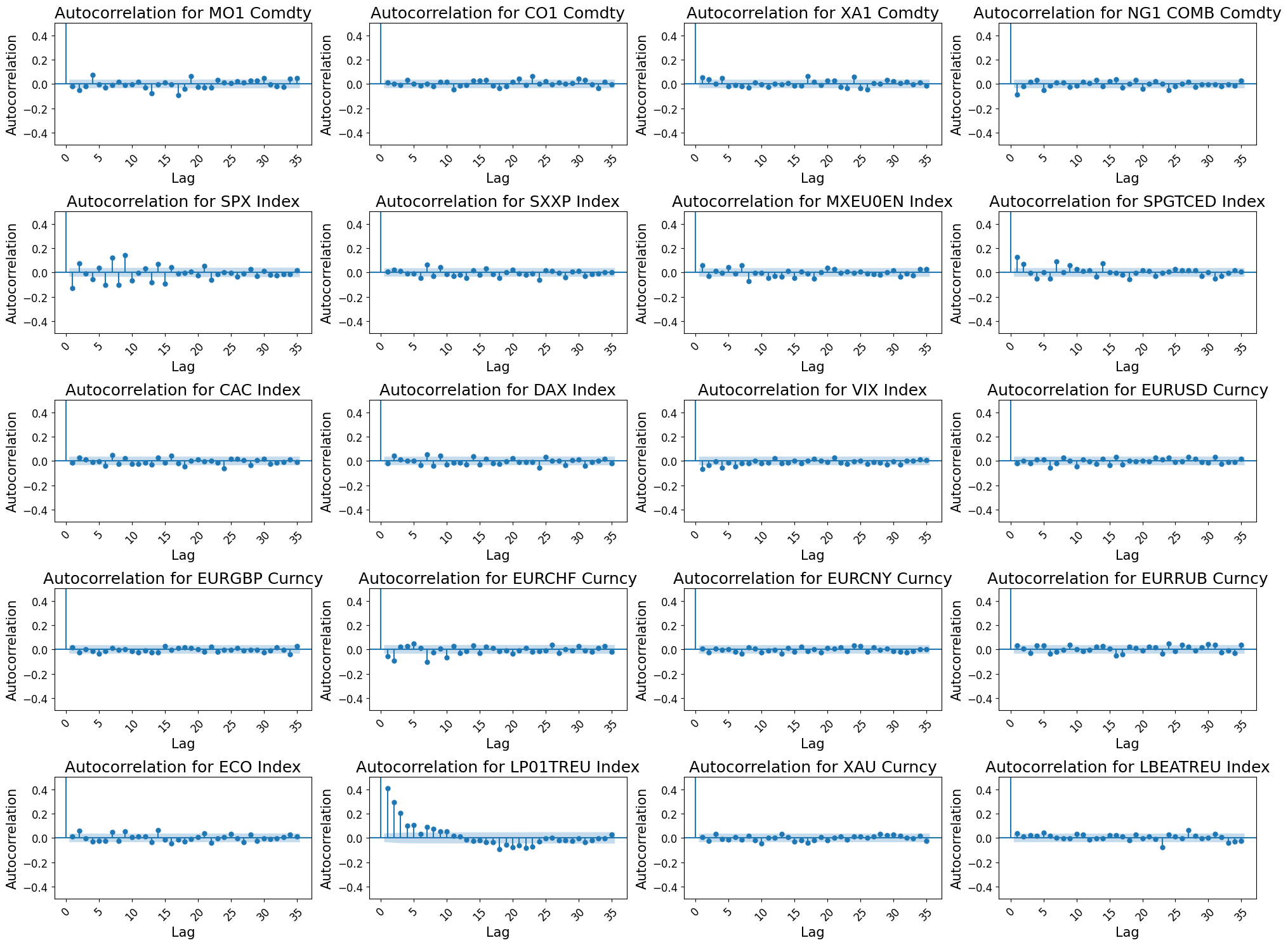}
\caption{Autocorrelation functions (ACF) of daily log returns for all variables included in the analysis. Shaded areas represent 95\% confidence intervals.}
\label{fig:acfplots}
\end{figure}

Autocorrelation is clearly present in most variables, with patterns varying across lags. As shown in Figure~\ref{fig:acfplots}, the autocorrelation functions for several series exceed the 95\% confidence bounds at multiple lags, indicating meaningful temporal dependence.

\begin{figure}
\centering
\includegraphics[width=0.85\textwidth]{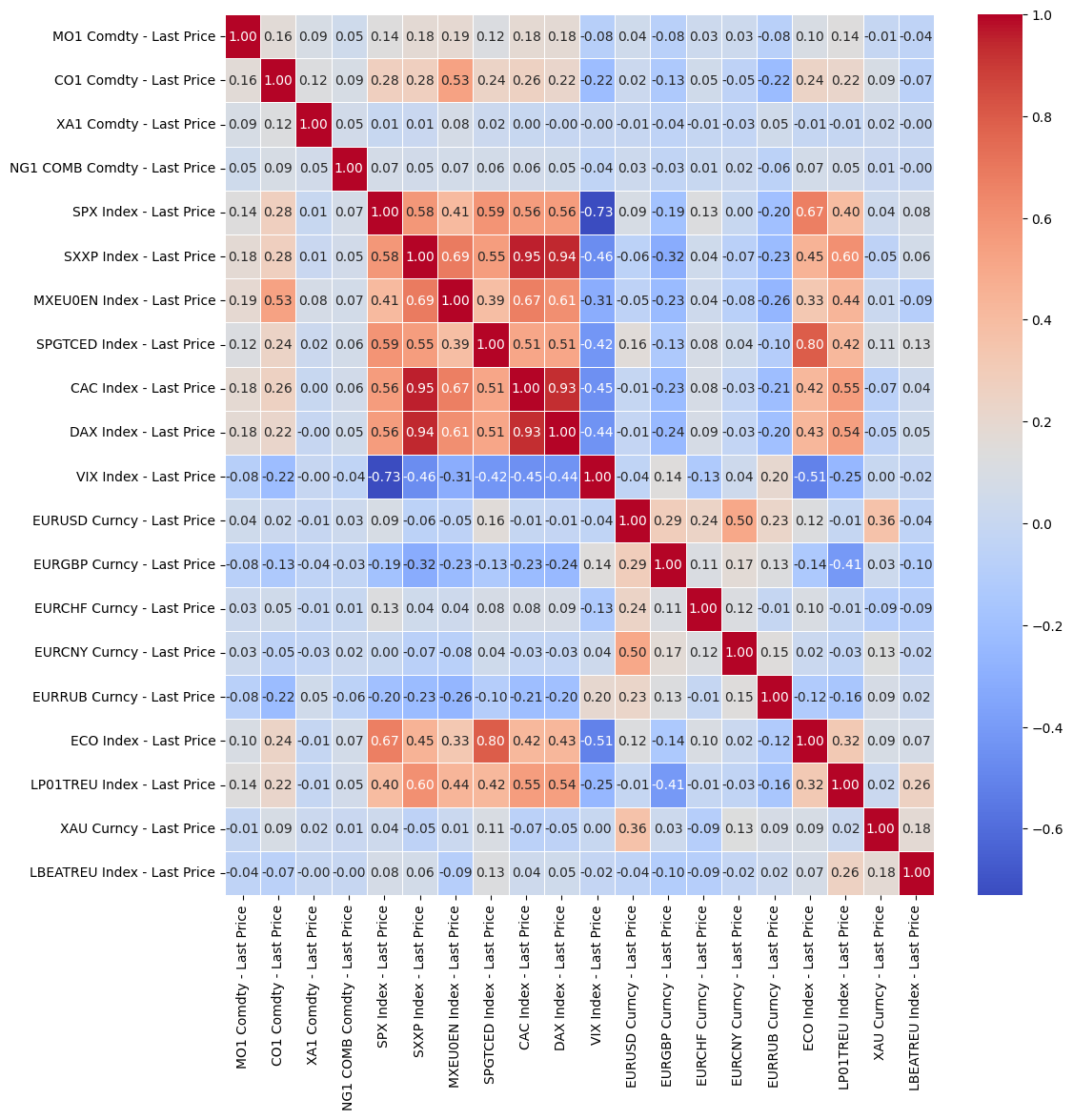}
\caption{Pearson correlation matrix of daily log returns for all variables in the analysis.}
\label{fig:correlationplot}
\end{figure}

Figure~\ref{fig:correlationplot} displays the correlation matrix of the log-return series. Strong positive correlations are observed among the stock indices and among the energy and clean energy equity indices, likely due to shared exposure to global macroeconomic conditions. These indices also exhibit high positive correlation with the LP01TREU Index, which tracks high-yield bonds and is highly sensitive to macro-financial risk. Conversely, the VIX index is negatively correlated with most stock indices, consistent with its role as a measure of market uncertainty and downside risk. Brent crude oil (CO1 Comdty) shows a notable positive correlation with the MSCI Europe Energy Index (MXEU0EN Index), reflecting its importance in energy production and pricing. Most remaining variables show low to moderate levels of correlation.

\newpage 

\section{AR-GARCH modelling}
\label{appendix:archgarch}

Before estimating the networks, we scaled all log-return series by a factor of 1,000 to avoid optimization issues associated with small values. For the static BN, we removed both autocorrelation and conditional heteroskedasticity by fitting AR-GARCH models to each series. Model orders for the autoregressive ($\text{lag}$) and GARCH components ($p$, $q$) were selected by minimizing the Bayesian Information Criterion (BIC), based on an exhaustive grid search over $\text{lag} \in {0, \dots, 7}$ and $p, q \in {1, \dots, 9}$. For the DBN, we retained autocorrelation to allow the model to capture temporal structure. Accordingly, we removed only heteroskedasticity using GARCH models (with AR order fixed to zero) and selected $p$ and $q$ by minimizing the BIC. After filtering, all standardized residuals were rescaled back to their original units by dividing by 1,000. The best-fitting model parameters for each series are summarized in Table~\ref{tab:combined_garch}.

\begin{table}[h]
\centering
\footnotesize
\renewcommand{\arraystretch}{1.3}
\scalebox{0.75}{
\begin{tabular}{lccc|cc|cc}
\toprule
\textbf{Variable} & \multicolumn{3}{c|}{\textbf{Static BN}} & \multicolumn{2}{c|}{\textbf{Dynamic BN}} & \textbf{BIC (BN)} & \textbf{BIC (DBN)} \\
& \textbf{Lag} & \textbf{$p$} & \textbf{$q$} & \textbf{$p$} & \textbf{$q$} & & \\
\midrule
CAC Index & 1 & 1 & 1 & 1 & 1 & 23032.91 & 23054.37 \\
CO1 Comdty & 2 & 1 & 1 & 1 & 1 & 27151.13 & 27147.70 \\
DAX Index & 1 & 1 & 1 & 1 & 1 & 23183.16 & 23207.69 \\
ECO Index & 5 & 1 & 1 & 1 & 1 & 27178.14 & 27187.84 \\
EURCHF Curncy & 2 & 1 & 1 & 1 & 1 & 15248.17 & 15256.10 \\
EURCNY Curncy & 2 & 1 & 1 & 1 & 1 & 17737.70 & 17733.17 \\
EURGBP Curncy & 2 & 1 & 1 & 1 & 1 & 17817.95 & 17813.44 \\
EURRUB Curncy & 7 & 1 & 2 & 1 & 2 & 22699.69 & 22701.91 \\
EURUSD Curncy & 7 & 1 & 1 & 1 & 1 & 18264.75 & 18266.72 \\
LBEATREU Index & 0 & 1 & 1 & 1 & 1 & 13373.94 & 13396.18 \\
LP01TREU Index & 2 & 1 & 1 & 1 & 1 & 11344.58 & 11859.61 \\
MO1 Comdty & 7 & 1 & 1 & 1 & 1 & 29216.37 & 29251.07 \\
MXEU0EN Index & 3 & 1 & 2 & 1 & 1 & 24764.02 & 24768.37 \\
NG1 COMB Comdty & 7 & 1 & 1 & 1 & 1 & 30311.70 & 30327.76 \\
SPGTCED Index & 7 & 1 & 1 & 1 & 1 & 24417.20 & 24482.17 \\
SPX Index & 1 & 1 & 1 & 1 & 1 & 21854.04 & 21915.32 \\
SXXP Index & 0 & 1 & 1 & 1 & 1 & 21979.08 & 22007.18 \\
VIX Index & 7 & 1 & 1 & 1 & 1 & 35046.67 & 35137.28 \\
XA1 Comdty & 4 & 1 & 1 & 1 & 1 & 23832.36 & 23855.78 \\
XAU Curncy & 3 & 1 & 1 & 1 & 1 & 22174.70 & 22173.20 \\
\bottomrule
\end{tabular}
}
\caption{Best-fitting GARCH model parameters for each variable. For the static BN, AR-GARCH$(\text{lag}, p, q)$ models were fitted; for the DBN, GARCH$(p, q)$ models were used with no AR component. BIC scores are reported for both.}
\label{tab:combined_garch}
\end{table}

\newpage
\section{Local sensitivity analysis: Tornado plots}
\label{appendix:tornado}

\begin{figure}[ht]
\centering
\includegraphics[width=0.75\textwidth]{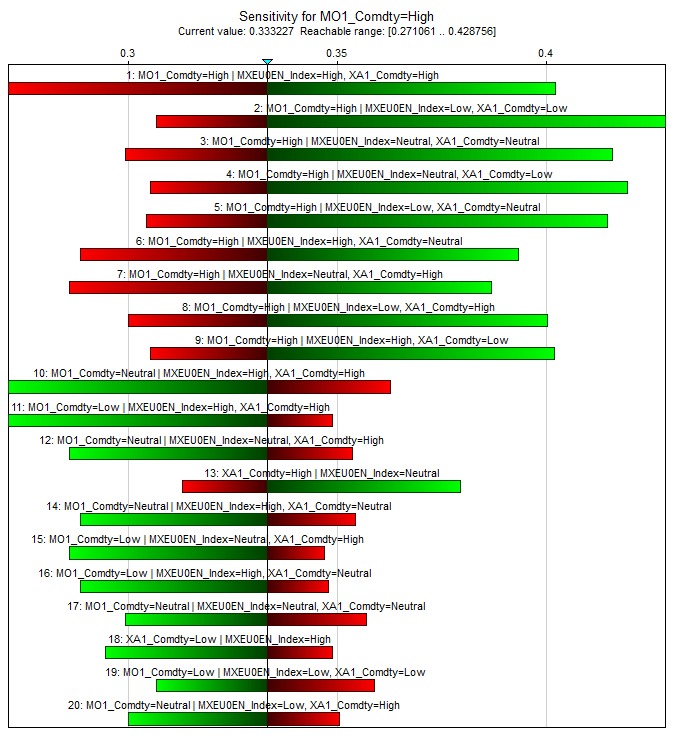}
\caption{Tornado plot for the high return outcome of EUA futures (MO1 = High).}
\label{fig:tornado_high_full}
\end{figure}

\begin{figure}
\centering
\includegraphics[width=0.75\textwidth]{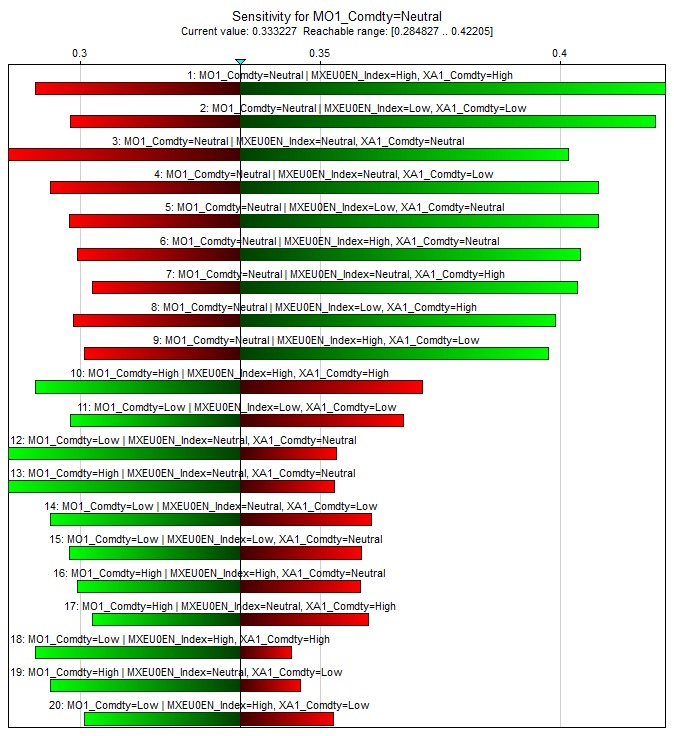}
\caption{Tornado plot for the neutral return outcome of EUA futures (MO1 = Neutral).}
\label{fig:tornado_neutral_full}
\end{figure}

\begin{figure}
\centering
\includegraphics[width=0.75\textwidth]{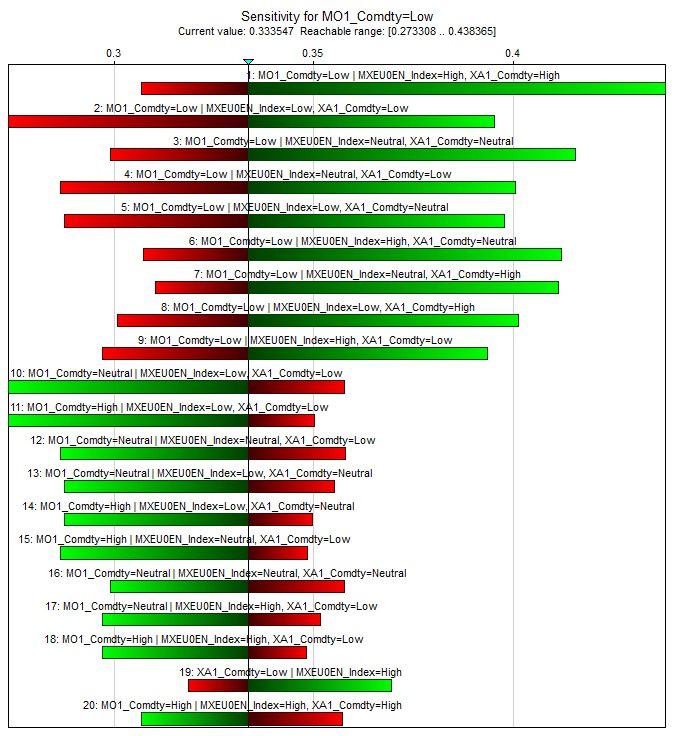}
\caption{Tornado plot for the low return outcome of EUA futures (MO1 = Low). }
\label{fig:tornado_low_full}
\end{figure}

\end{document}